\documentclass[12pt,a4paper]{article} 

\usepackage{ams}
\usepackage{amsmath}
\usepackage{amsthm}
\usepackage[english]{babel}
\usepackage{graphicx}
\usepackage{anysize}
\usepackage{amsfonts} 
\usepackage{amssymb} 
\usepackage{mathtools}
\usepackage{booktabs}
\usepackage{dsfont}
\usepackage{hyperref} 
\usepackage{cite}
\usepackage{float}

\newcommand{\pderiv}[1]{\frac{\partial}{\partial #1}}

\newcommand{\oppr}[3]{\ensuremath{\left \langle \left. #1\right.  \right| #2 \left| \left. #3 \right. \right \rangle}}
\newcommand{\ket}[1] {\ensuremath {\left| #1 \right\rangle }}
\newcommand{\bra}[1] {\ensuremath {\left \langle #1 \right|}}
\newcommand{\inpr}[2] {\ensuremath {\left \langle \left. #1 \right| #2 \right\rangle}}
\newcommand{\outpr}[2] {\ensuremath {\left | \left. #1 \right \rangle \right \langle #2|}}

\newcommand{\overroottwo}[1] {\frac{#1}{\sqrt{2}}}
\newcommand{\twomatrix}[4] {\ensuremath{\left[\begin{array}{cc} #1 & #2 \\ #3 & #4\end{array}\right]}}
\newcommand{\twovector}[2] {\ensuremath{\left(\begin{array}{c} #1 \\ #2 \end{array}\right)}}

\newcommand{\fourvector}[4] {\ensuremath{\left(\begin{array}{c} #1 \\ #2 \\ #3 \\ #4 \end{array}\right)}}

\marginsize{3.2cm}{3.2cm}{2cm}{2cm}

\title{\textsc{Quantum Computing in the de Broglie-Bohm Pilot-Wave Picture}} 
\author{Philipp Roser\footnote{philipp.roser09@imperial.ac.uk}}
\date{September 2010}

\begin{document}

%_________________________________________________________________________________

\begin{titlepage}
	\maketitle
	\thispagestyle{empty}
	\center{
	Blackett Laboratory \\Imperial College London\\
	\vspace{10mm}
	Submitted in partial fulfilment of the requirements for the degree of\\ Master of Science of Imperial College London\\
	\vspace{10mm} 	
	Supervisor: Dr.\ Antony Valentini\\ Internal Supervisor: Prof.\ Jonathan Halliwell\\
	}
	
	\vspace{10mm}	
	\begin{abstract}

		Much attention has been drawn to quantum computing and the exponential speed-up in computation the technology would be able to provide. Various claims have been made about what aspect of quantum mechanics causes this speed-up. Formulations of quantum computing have traditionally been made in orthodox (Copenhagen) and sometimes many-worlds quantum mechanics. We will aim to understand quantum computing in terms of de Broglie-Bohm Pilot-Wave Theory by considering different simple systems that may function as a basic quantum computer. We will provide a careful discussion of Pilot-Wave Theory and evaluate criticisms of the theory. We will assess claims regarding what causes the exponential speed-up in the light of our analysis and the fact that Pilot-Wave Theory is perfectly able to account for the phenomena involved in quantum computing.
		
	\end{abstract}
\end{titlepage}

\baselineskip = 24pt

I, Philipp Roser, hereby confirm that this dissertation is entirely my own work. Where other sources have been used, these have been clearly referenced.

\pagebreak
\tableofcontents
\pagebreak

%_________________________________________________________________________________

	\section{Introduction}
			
		A large number of books about quantum mechanics begin by citing the theory's success in accounting for non-classical phenomena that were discovered in the early parts of the twentieth century. Many such books then continue by stating that the reason we believe in the theory of quantum mechanics today is not mathematical beauty or simplicity, but the fact that it works and does so to an incredible degree of accuracy. That is, it ``works'' in that its predictions agree phenomenally well with observed experimental outcomes.
		
	There appears to be a broad consensus in the physics community that quantum mechanics (with quantum field theory etc.) is a good theory because it has been so well corroborated experimentally. Some physicists may be willing to go further and say that it follows that there is something \emph{true} about quantum mechanics, that is, that nature in some sense really behaves as the theory describes, with collapsing wave functions and spooky action at a distance. Others may take the positivist view along with Bohr and claim that one cannot meaningfully talk about what nature is really like but only about the outcomes of measurements. Others still may take no such stand at all and accept quantum mechanics simply as a powerful toolbox with which to make predictions about experiments, abstaining from any ``interpretational'' questions. Yet all these positions base their faith in quantum mechanics on the high agreement between theory and experiment.
	
	It would then seem reasonable to suppose that any other theory that is equally well corroborated (and internally consistent) should find similar support. Indeed, one might be led to believe that any other internally consistent theory that makes exactly the same predictions as standard quantum theory should be treated on par with it. As a matter of fact, there is a theory that provably makes the same predictions as standard quantum theory, namely what has become misleadingly known as ``de Broglie-Bohm'' Pilot-Wave Theory, or sometimes, even more misleadingly, ``Bohmian'' mechanics.\footnote{As we will discuss in section \ref{PWT}, de Broglie's and Bohm's theories are not exactly identical.} However, among physicists Pilot-Wave Theory (PWT) is often either misunderstood or, more frequently, simply unknown. Thus it is not treated on par with standard quantum mechanics, not because of malicious intent to suppress unorthodox views (or any similar such conspiratorial reason), but because of mere unfamiliarity. There have been various attempts to communicate the feasibility of PWT to a wider audience, notably through the works of Bell\footnote{1987, \cite{Bell1987}}, Bohm and Hiley\footnote{1993, \cite{BohmHiley1993}}, Holland\footnote{1993,\cite{Holland1993}} and others. While in the last twenty years or so as a result of these works PWT has become more accepted as an alternative formulation of quantum mechanics and recently has received some attention for example in arrival time calculations (a concept with which standard quantum mechanics struggles), as a whole it is still not nearly as well known as it ought to be. One purpose of this work is to provide a clear and succinct presentation of PWT and thereby hopefully help to promote it as a feasible theory that accounts for experimental outcomes just as well as standard quantum mechanics does.
	
	One may seek the reason why PWT never became as widely known, well understood, or popular as standard quantum mechanics in history. De Broglie presented a preliminary (yet incomplete) account of his theory in his PhD thesis in 1924, published in 1925 in \emph{Annales de Physique},\footnote{PhD Thesis, 1925,\cite{deBrogliePhD}} and a complete version in May 1927, which formed his presentation at the Solvay Conference in October of the same year\footnote{1927,\cite{deBroglie1927}}. Contrary to popular belief it also contained an account of PWT for many-particle systems. Perhaps the main reason the quantum mechanics of Bohr and Heisenberg became more widely accepted than de Broglie's Pilot-Wave Theory was an objection by Pauli regarding inelastic scattering. %\footnote{As part of his criticism Pauli is said to have come up with the subsequently famous line that pilot-wave theory was ``not even wrong'' (\emph{ganz falsch}), which today has}. 
 In fact, de Broglie had the machinery to deal with the objection, but historically there has been much confusion over the precise content of the discussion, not least due to Bohm's misrepresentation of the objection.\footnote{1952, appendix B,\cite{Bohm1952}} Furthermore, de Broglie abandoned his theory a few years after the Solvay Conference for other reasons. For a detailed account of the conference and de Broglie's papers see Bacciagaluppi and Valentini.\footnote{2009 (chapters 2, 10, 11) \cite{BacciagaluppiValentini2009}} 
	
	Bohm rediscovered the theory (with minor differences), including a detailed description of measurement processes, in 1952, around the time he was forced to leave Princeton for Brazil for suspected communist sympathies. McCarthyism, together with the then established traditional quantum mechanics of Bohr, meant that Bohm's work received little to no attention. Oppenheimer even suggested, ``If we cannot disprove Bohm, we must agree to ignore him.'' Until the 1990s, with the notable exception of Bell, little attention had been drawn to either de Broglie's or Bohm's work. Furthermore, in the decades following the publication of his 1952 papers, Bohm's ideas and interest in more metaphysical questions have alienated many physicists from his work in general.\footnote{And the fact that Bohm's ideas have made him strangely popular with various ``New Age'' and spiritual movements has not helped his professional reputation either.} 
	
	Leaving historical issues aside, one may argue that standard quantum theory and PWT are not to be treated on par for reasons that are not to do with empirical results but internal consistency. There have been a variety of objections to PWT ranging from the ill-informed to the apparently reasonable. I will address some of these in section \ref{PWT} and the appendix. A large number of possible objections are summarised in Mike Towler's lecture notes on PWT.\footnote{Pilot-wave theory, Bohmian metaphysics, and the foundations of quantum mechanic (2009), \cite{TowlerLectures}}
	
	In fact, PWT still does have some genuinely open questions. However, one of its great virtues is a coherent account of the measurement process, as well as a consistent theory of quantum cosmology. PWT does not suffer from the infamous ``measurement problem'' of standard quantum mechanics (SQM).\footnote{One ought to strictly differentiate between Copenhagen quantum mechanics, postulating a physical collapse of the wave function, and an instrumentalist account, taking the attitude that only predictions and measurement outcomes matter. Both might be seen as being somewhat ``standard'' version of quantum mechanics. Both have a measurement problem (even though to instrumentalists its existence is irrelevant) and for our purposes it will suffice to consider them together as ``SQM''.} This is not the place for a detailed discussion of the measurement problem and the various attempts to overcome it, although some comments will be made in later sections about the Everettian ``many-worlds'' interpretation in the context of claims by David Deutsch.\footnote{For a detailed review article of the measurement problem, including a discussion of the importance of decoherence, see Wallace, \emph{Philosophy of Quantum Mechanics}, in Rickles, 2008, \cite{Rickles2008}} In brief, PWT does not rely on an \emph{ad hoc} distinction between the quantum system and a (semi-)classical apparatus/environment but, just like in the Everettian theory, treats the measurement apparatus just like any other physical system with which the ``measured'' system interacts. However, one ought to draw attention to the fact that PWT is \emph{not} a theory specifically constructed to overcome the measurement problem, but was conceived by de Broglie as a response to unexplained empirical phenomena. This is important to remember when confronted with criticisms of PWT suggesting that it is a contrived theory constructed only for the purpose of escaping the measurement problem.
	
	We should not reject PWT at the sight of a first ontological or other difficulty, given that for eighty years physicists have used SQM despite its serious and still unresolved problems. If nothing else, PWT functions as an \emph{effective} theory, just as SQM functions as one of the most successful effective theories of all time (perhaps only rivalled by thermodynamics), but as a fundamental theory is either incomplete or incorrect.
	
	The lack of familiarity of the physics community with PWT has meant that, while proof that the theory reproduces the results of standard quantum mechanics exactly is easy to obtain, phenomena concerning a number of applications of quantum mechanics have not yet ever been described in detail in the PWT picture. To our knowledge, this includes a description of quantum computing. To provide such a description, at least for a particular simple quantum algorithm, is another purpose of this work.
	
	As indicated already we will see that PWT reproduces the experimental results of standard quantum mechanics for all physical systems exactly. Hence there is no question that quantum computers work in a world that is fundamentally governed by pilot-wave dynamics. No matter how a quantum computer is constructed, it is some physical system at the end of which there must be some measurement that constitutes the result of running a particular algorithm. Thus this result is the same for orthodox quantum theory and PWT. Relevant aspects of (standard) quantum computing are reviewed in section \ref{QC}. Section \ref{QCinPWT} constitutes an analysis of the so-called Deutsch algorithm in the PWT picture. Since this requires a dynamical description of the computer, we will look at two particular possible (though not necessarily realistic) implementations of two-qubit systems: in terms of energy states of a particle in an infinite well and in terms of spin-$\frac{1}{2}$ particles.
	
	The purpose is not merely pedagogical. Various claims have been made about what allows for the exponential speed-up of quantum computing as compared to classical computing. While one may argue that the question itself is vague or even misguided, we will examine some of these claims and counterclaims (most notably by David Deutsch,\footnote{1985b,\cite{Deutsch1985b} and 1997, \cite{Deutsch1997}} Richard Josza,\footnote{in Huggett (1998) \cite{Huggett1998}}, David Mermin\footnote{2007a\&b, \cite{Mermin2007a, Mermin2007b}} and Andrew Steane\footnote{2002, \cite{Steane2002}}) in the context of the possibility of a PWT quantum computer, followed by some concluding remarks.

%______________________________________________________________________________________________________________________		
		
	\pagebreak		
	\section{De Broglie-Bohm Pilot-Wave Theory}\label{PWT}
			
	\subsection{Motivation}
	
	At the heart of any physical theory is the need to account for observed data. During parts of the twentieth century (the era of positivism) the predominant view was that this is \emph{all} a theory should do. We have seen that SQM as conceived by Bohr and Heisenberg readily accounts for a great variety of microscopic phenomena. However, one has good reason to doubt the \emph{explanatory value} of a theory whose basic postulates include a notion called ``measurement'' that escapes rigorous definition within the context of the theory. Talking of ``explanatory value'' carries a long tail of philosophical debate in which we will not engage here, and in fact we do not have to. The problem of measurement in SQM does not merely detract from its explanatory power, but it may even lead us to question the consistency of the theory: Experimental apparatuses and experimenters are both made of atoms obeying the deterministic law of evolution encoded in the Schr\"odinger equation (in non-relativistic SQM), but the experimenters and apparatuses themselves appear as fundamental entities in the infamous ``collapse postulate''. Not only do they have some \emph{holistic} dynamical effect that cannot be accounted for by the dynamics of their constituent parts, but neither is it clear what exactly constitutes a ``measurement'' or an ``experimenter''. SQM nevertheless functions as an impressive effective theory.
	
	PWT was conceived in order to account for observed data, not in response to the measurement problem, which only became understood many years later. However, today the measurement problem might well be motivation to consider PWT as a serious competitor of SQM, although it is misleading to say that PWT \emph{solves} the measurement problem. Rather, it is an alternative theory that simply \emph{does not have} a measurement problem. 
	
	It is sometimes argued that one of the advantages of PWT over SQM is that it is fundamentally deterministic, even if we can never actually obtain knowledge of the future in the same way we can in Newtonian physics. I disagree with this point of view. There is no \emph{a priori} reason why the world should be deterministic at a fundamental level.\footnote{...if not for controversial metaphysical or even theological reasons. While we should pay attention to good, cogent arguments from analytical philosophy, in this case we do not know of any such argument.} However, the deterministic nature of PWT does serve to show that a deterministic explanation of the outcomes quantum experiments is \emph{possible}, contrary to widely held opinions found both in the physics community as well as the portrayal of quantum theory in popular literature.
	
	PWT is a theory with real particle trajectories. While proponents of the theory might cite this fact as making the dynamics precise, or definite, opponents have previously condemned the theory for it. However, why should we view the existence of trajectories as either positive or negative? There are no fundamental principles dictating that theories should (or should not) have trajectories. In any case, the trajectories of PWT are not \emph{classical} trajectories (so the theory is definitely not ``too classical'') but obey highly non-classical dynamics and they are themselves unobservable. Unfortunately Bohm's 1952\footnote{1952a,b \cite{Bohm1952}} version of PWT is framed in a way that gives it the appearance of a modified classical theory, which has confused matters. PWT is an independent theory and how far it is similar to other classical or quantum theories is irrelevant in this context. What is important is that PWT is a consistent theory that accounts for experimental data.\\

	\subsection{Two theories of pilot-waves}
	
	PWT was developed by de Broglie during the 1920s but received little approval at the famous Solvay conference of 1927.\footnote{see Bacciagaluppi and Valentini, 2009, \cite{BacciagaluppiValentini2009}} De Broglie formulated the theory in terms of a first-order mechanics, relating the velocity of a ``particle'' in configuration space to the phase of the pilot-wave, which itself obeys the Schr\"odinger equation. Bohm's formulation is of second order and obtained by differentiating de Broglie's equation. The law of motion is often rewritten in the form a Hamilton-Jacobi equation with an additional ``quantum potential'' term that is a function of (the second derivative of) the wave function, giving the theory a pseudo-classical appearance. The differentiation implies that there is an extra integration constant to be fixed in the calculation of trajectories. Hence, for initial values disagreeing with de Broglie's formulation, a wider physics may be obtained. In this sense, we should strictly speak of two different pilot-wave theories, where one reduces to the other for a specific choice of integration constant. However, this wider physics leads to experimental outcomes that disagree with actual outcomes and so we are naturally interested in the special choice of constant that agrees with de Broglie's theory. We will not consider Bohm's wider physics here any further and use the term ``Pilot-Wave Theory'' (PWT) interchangably, except where otherwise indicated.
	
	%\subsubsection{Fundamentals of PWT in de Broglie's theory}
	
	We will now present the formalism of non-relativistic PWT. We acknowledge that most readers will be more familiar with orthodox quantum mechanics and so, to aid understanding, corresponding ideas in SQM will occasionally be referred to. Note though that under no circumstances should it be assumed that any of the principles of SQM apply, unless stated otherwise. We use units such that $\hbar=1$ and $c=1$ throughout.
	
	PWT proposes the existence of two entities, a pilot-wave and a ``particle'' (also variously referred to as ``corpuscle''\footnote{e.g.\ in Brown \& Wallace, 2007, \cite{BrownWallace2007}} or simply ``configuration''). There remains some controversy over the ontological status of the pilot-wave, which we will discuss in section \ref{objections}. The complex pilot-wave function $\psi(\vec{x},t)$ of a system $\mathfrak{S}$ obeys the equation
	\begin{eqnarray} i\frac{\partial}{\partial t}\psi(\vec{x},t) = \hat{H}_{\mathfrak{S}}\psi(\vec{x},t) \end{eqnarray}
where $\hat{H}_{\mathfrak{S}}$ refers to some operator encoding the Hamiltonian of the system. Note that operators here are merely mathematical objects that are not associated with any form of ``observable''.

	For many physical systems the equation takes the form
	\begin{eqnarray}\label{SE} i\frac{\partial}{\partial t}\psi(\vec{x},t) = \sum_{i=1}^{N}{-\frac{1}{2m_i}\nabla_i^2\psi(\vec{x},t)}+V(\vec{x},t)\psi(\vec{x},t) \end{eqnarray}
where $V(\vec{x},t)$ is a potential, $i$ labels the particles and $\nabla_i$ is the three-dimensional spatial derivative in the $i$th particle's coordinates. 

	Although we refer to this equation as the ``Schr\"odinger equation'', there is a fundamental difference between this equation and the Schr\"odinger equation of SQM. In the latter, the equation describes the evolution of a state in \emph{Hilbert space}, which in many cases can be expressed in the position basis by taking the inner product of the state with \ket{x}. The pilot-wave on the other hand exists in $3N$-dimensional \emph{configuration space} $\mathds{R}^{3N}$, where $N$ is the number of physical particles in $\mathfrak{S}$. Thus the wave equation of PWT looks always like the SQM Schr\"odinger equation in the position basis.

	The particle is specified entirely through a $3N$-dimensional vector $\vec{x}$ denoting its position in configuration space. Thus $\vec{x}$ encodes all degrees of freedom of $\mathfrak{S}$. Perhaps more accurately we should say that the ``particle'' just \emph{is} the position (i.e.\ the configuration) specified by $\vec{x}$. It is not a particle comparable to physical particles in 3-space. To think of a moving point in configuration space is variously helpful or misleading. Note that PWT requires the reality of configuration space.\footnote{A version of PWT without configuration space in its ontology may be conceivable, but requires an apparently arbitrary division of what is ``real''. See section \ref{objections}.} There is yet an open question about the ontology of 3-space, namely whether 3-space is supposed to supervene on configuration space or whether is a separate entity. Neither answer seems entirely satisfactory. The problem and how it reflects on PWT is discussed in section \ref{objections}, too.
	
	The particle has at all times a definite position and velocity in configuration space and so the physical particles of $\mathfrak{S}$ have a definite position and velocity at all times. How to recover the probabilistic nature of the predictions of SQM will be described in section \ref{ensemble}. In de Broglie's formulation, we postulate that for a general Hamiltonian $H_{\mathfrak{S}}$ the equation governing the motion of the particle in configuration space is given by the guidance equation
	\begin{eqnarray}\label{guidanceequation} \dot{\vec{x}}(t) = \frac{\vec{j}(\vec{x},t)}{|\psi(\vec{x},t)|^2} \end{eqnarray}
where $\vec{j}(\vec{x},t)$ is a current implicitly defined as the current appearing in the continuity equation
	\begin{eqnarray}\label{current}\frac{\partial}{\partial t}|\psi(\vec{x},t)|^2 + \nabla\cdot\vec{j}(\vec{x},t) = 0, \end{eqnarray} 
which is derived from the Schr\"odinger equation. We note that there is an apparent gauge\footnote{We should not take this label (``gauge'') too literally: While it can be shown that different choices of $\vec{a}$ do not alter any empirical predictions, they do differ on the actual paths of the particles in space.} freedom $\vec{j}(\vec{x},t)\rightarrow\vec{j}(\vec{x},t)+\vec{a}(\vec{x},t)$ such that $\nabla\cdot\vec{a}(\vec{x},t)=0$. We will return to this in section \ref{objections}.
From the Schr\"odinger equation and its Hermitian conjugate we derive
\begin{eqnarray}\notag i\frac{\partial}{\partial t}|\psi(\vec{x},t)|^2 
&=& i\left(\frac{\partial}{\partial t}\psi^*(\vec{x},t)\right)\psi(\vec{x},t)
			 +i\psi^*(\vec{x},t)\left(\frac{\partial}{\partial t}\psi(\vec{x},t)\right)  \\
&=& -\psi(\vec{x},t)\hat{H}^{\dagger}_{\mathfrak{S}}\psi^*(\vec{x},t)+\psi^*(\vec{x},t)\hat{H}_{\mathfrak{S}}\psi(\vec{x},t)
\end{eqnarray}
and so
\begin{eqnarray}
\frac{\partial}{\partial t}|\psi(\vec{x},t)|^2 = i\left(\psi(\vec{x},t)\hat{H}^{\dagger}_{\mathfrak{S}}\psi^*(\vec{x},t)-\psi^*(\vec{x},t)\hat{H}_{\mathfrak{S}}\psi(\vec{x},t)\right).
\end{eqnarray}
Recall that the operator $\hat{H}_{\mathfrak{S}}$ is a mathematical object only and not explicitly associated with an observable or any such concept. Hermiticity of $\hat{H}_{\mathfrak{S}}$ is of no particular physical significance.

	For the case 
\begin{eqnarray} \hat{H}_{\mathfrak{S}}\psi(\vec{x},t) = \sum_{i=1}^{N}{-\frac{1}{2m_i}\nabla_i^2\psi(\vec{x},t)}+V(\vec{x},t)\psi(\vec{x},t) \end{eqnarray}
we find
\begin{eqnarray} \frac{\partial}{\partial t}|\psi(\vec{x},t)|^2 = 
\sum_{i=1}^{N}{\frac{i}{2m_i}(\psi(\vec{x},t)\nabla_i^2\psi^*(\vec{x},t)-\psi^*(\vec{x},t)\nabla_i^2\psi(\vec{x},t))} \end{eqnarray}
and hence, from the definition of $\vec{j}(\vec{x},t)$ through its divergence,
\begin{eqnarray}\notag \vec{j_i}(\vec{x},t) 
&=& \frac{i}{2m_i}(\psi(\vec{x},t)\nabla_i\psi^*(\vec{x},t)-\psi^*(\vec{x},t)\nabla_i\psi(\vec{x},t)) \\
&=& \frac{1}{m_i}\mathfrak{Im}(\psi(\vec{x},t)\nabla_i\psi^*(\vec{x},t))  \end{eqnarray}
where $\vec{j_i}$ is the projection of $\vec{j}$ onto the three-dimensional space corresponding to the degrees of freedom of the $i$th particle.

	Writing $\psi(\vec{x},t)=R(\vec{x},t)e^{iS(\vec{x},t)}$, the velocity of the $i$th particle simplifies to
\begin{eqnarray}\notag \dot{\vec{x}}_i(t) 
&=& \frac{1}{R^2}\frac{i}{2m_i}(R^2(-i\nabla_iS)+R\nabla_iR-R^2(+i\nabla_iS)-R\nabla_iR) \\
&=&\label{deBroglieLoM} \frac{1}{m_i}\nabla_iS(\vec{x},t). \end{eqnarray}

	In SQM, $\vec{j}(\vec{x},t)$ would be interpreted as the probability current density vector. Its equation is a differential equation for the position $\vec{x}$ of the particle. 
	
	Equivalently we could have defined a velocity field $v^{\psi}$ over all of configuration space by
	\begin{eqnarray} v^{\psi}(\vec{x},t) = \frac{\vec{j}(\vec{x},t)}{|\psi(\vec{x},t)|^2}, \end{eqnarray}
together with the law for the particle, simply stated as
	\begin{eqnarray} \frac{d}{dt}{\vec{x}}(t) = v^{\psi}(\vec{x},t). \end{eqnarray}
	
	%\subsubsection{Bohm's theory of PWT}
	
	Bohm's formulation may be obtained by differentiating de Broglie's equation with respect to time, or directly from the Schr\"odinger equation. We will pursue the latter method here. Substituting $\psi(\vec{x},t)=R(\vec{x},t)e^{iS(\vec{x},t)}$ into the Schr\"odinger equation (\ref{SE}) yields
\begin{eqnarray}\notag i(\pderiv{t}R+Ri\pderiv{t}S)e^{iS}
&=&\notag \sum_{i=1}^N{-\frac{1}{2m_i}\nabla_i\cdot[(\nabla_iR+iR\nabla_iS)e^{iS}]} + VRe^{iS} \\
&=&\notag  \sum_{i=1}^N -\frac{1}{2m_i}[\nabla_i^2R+i\nabla_iR\cdot\nabla_iS   \\ 
&&\notag \qquad	+iR\nabla_i^2S+i(\nabla_iR+iR\nabla_iS)\cdot\nabla_iS]e^{iS} + VRe^{iS} \\ &&
\end{eqnarray} 
and hence
\begin{eqnarray} \notag i\frac{\partial R}{\partial t}-R\frac{\partial S}{\partial t} &=& \sum_{i=1}^N{-\frac{1}{2m_i}[\nabla_i^2R+2i\nabla_iR\cdot\nabla_i{S}+iR\nabla_i^2S-R(\nabla_iS)^2]}+VR. \\ \label{SEsplit}&&	\end{eqnarray}
Considering only the \emph{real} part gives
\begin{eqnarray} -R\frac{\partial S}{\partial t} = \sum_{i=1}^N-\frac{1}{2m_i}[\nabla_i^2R-R(\vec{\nabla}_iS)^2]+VR. \end{eqnarray}
Dividing by $R$ we obtain
\begin{eqnarray}\label{QHJ} -\frac{\partial S}{\partial t} = \sum_{i=1}^N\frac{(\nabla_iS)^2}{2m_i}+V+Q \end{eqnarray}
where 
\begin{eqnarray} Q = \sum_{i=1}^N-\frac{1}{2m_i}\frac{\nabla_i^2R}{R} \end{eqnarray}
is known as the quantum potential. Elsewhere (including Bohm's 1952 papers\footnote{1952a,b, \cite{Bohm1952}}) the quantum potential is denoted by $U$, but we will use $Q$ exclusively.

	Motivation to call $Q$ a ``potential'' arises from recognising equation \ref{QHJ} as a modified form of the Hamilton-Jacobi equation. Its standard form is recovered for $Q\rightarrow 0$. $Q$ and $V$ appear in the same capacity in the equation, hence $Q$ has, at least mathematically, the role of an additional potential term. To explicitly derive the corresponding equation of motion, apply $\nabla_j$ to equation \ref{QHJ} to find
\begin{eqnarray}\notag -\nabla_j(V+Q) &=& \nabla_j\frac{\partial S}{\partial t} + \nabla_j\sum_{i=1}^N\frac{1}{2m_i}(\nabla_iS\cdot\nabla_iS) \\
 &=& \frac{\partial}{\partial t}\nabla_jS + \sum_{i=1}^N\frac{1}{m_i}(\nabla_iS\cdot\nabla_j)\nabla_iS.  \end{eqnarray}
Using de Broglie's equation of motion (\ref{deBroglieLoM}) twice yields
\begin{eqnarray} -\nabla_j(V+Q) &=& \frac{\partial}{\partial t}(m_j\dot{\vec{x}}_j) + \sum_{i=1}^N m_i(\dot{\vec{x}}_i\cdot\nabla_j)\dot{\vec{x}}_i. \end{eqnarray}
Note now that the only non-zero contribution to the sum comes from the case $i=j$. Using $\frac{d}{dt} = \frac{\partial}{\partial t}+\dot{\vec{x}}\cdot\nabla$, Bohm's law of motion for the $j$th particle is 
\begin{eqnarray} m_j \ddot{\vec{x}}_j = -\nabla_j(V+Q).  \end{eqnarray}
This is a second-order equation, which when integrated yields de Broglie's equation up to a constant. Thus, if this constant is not fixed by some boundary condition (such as de Broglie's equation for some time $t_0$, Bohm's dynamics in principle allows a wider physics than de Broglie's.

	According to Bohm, ``[t]he equation for $S$ implies, however, that the particles move under the action of a force which is not entirely derivable from the classical potential ... but which also obtains a contribution from the `quantum mechanical' potential.''\footnote{1952a, p. 170, \cite{Bohm1952}} This pedagogy is misleading. $Q$ is only ``like $V$'' in that it adds linearly to $V$ in the modified Hamilton-Jacobi equation. Physically however it is a fundamentally different kind of entity. While $V$ is determined by the physical environment such as nearby charges or massive bodies, $Q$ is a function of the pilot-wave function, which \emph{evolves in time}. 
	
	Classically, the potential $V$ is a useful concept since the total energy, i.e.\ the sum of potential and kinetic energy, is always conserved. This is not the case if $Q$ is added to the classical potential. The reason we consider the concept of ``energy'' at all is purely \emph{because it is conserved}. Furthermore, to speak of a ``force'' containing a contribution from the quantum potential deprives the concept of force of its physical meaning and reduces it to be defined purely as ``whatever is on the right hand side of $m\ddot{\vec{x}} = \ldots$''. The view thereby imposes the classical equations of motion to hold no matter what, even if mysterious quantum forces have to be introduced. Yet the theory is fundamentally non-classical. Why not simply accept that quantum systems do not obey classical dynamics, but a completely different dynamics as, for example, given by de Broglie's formulation?
	
	An advantage of using Bohm's pseudo-Hamilton-Jacobi formalism is that a lot of mathematical machinery has already been developed that may be employed here, but to speak of $Q$ as being a physical potential seems unmotivated. We will treat Bohm's formulation of PWT as a purely mathematical toolbox, while taking de Broglie's treatment as the more fundamental.
	
	The modified Hamilton-Jacobi equation was derived by considering the real part of equation \ref{SEsplit}. Taking now the imaginary part gives
	\begin{eqnarray} \frac{\partial R}{\partial t} = \sum_{i=1}^N[-\frac{1}{m}\nabla_iR\nabla_iS - \frac{1}{2m}R\nabla_i^2S] \end{eqnarray}
and hence after multiplication by $2R$
	\begin{eqnarray}\notag 0 &=& 2R\frac{\partial R}{\partial t} + \sum_{i=1}^N[\frac{2R}{m_i}\nabla_iR\nabla_iS + \frac{R^2}{m_i}\nabla_i^2S] \\
&=&\notag \frac{\partial R^2}{\partial t} + \sum_{i=1}^N \nabla_i \cdot \left(\frac{R^2\nabla_iS}{m_i}\right) \\
&=& \label{ImaginarySE} \frac{\partial R^2}{\partial t} + \sum_{i=1}^N \nabla_i \cdot (R^2 \dot{\vec{x}}_i), \end{eqnarray}
where we have used de Broglie's equation (\ref{deBroglieLoM}) in the last line. We will return to this equation below. For the single particle case the sum and the corresponding indices $i$ may be dropped. 

We will now proceed with two simple examples.
	
	%\subsubsection{Examples}
	
	\textbf{The plane wave.} For $V=0$ we obtain a plane wave solution for the Schr\"odinger equation (\ref{SE}):
\begin{eqnarray} \psi(\vec{x},t) = Ae^{i(\vec{p}\cdot\vec{x}-Et)} \end{eqnarray}
where $\vec{p}$ is the momentum of the wave, E the associated energy and $A$ a normalisation constant. Then for a single particle of mass $m$ de Broglie's equation of motion is
\begin{eqnarray} m\dot{\vec{x}} = \nabla S = \nabla(\vec{p}\cdot\vec{x}-Et) = \vec{p}. \end{eqnarray}
Hence
\begin{eqnarray} \vec{x}(t) = \vec{x}_0+\frac{\vec{p}}{m}t, \end{eqnarray}
i.e.\ the particle moves at a constant velocity equal to the wave momentum divided by the particle's mass. This agrees with the classical Newtonian result, although note that for the vast majority of systems there is no such agreement with classical physics. 
	
	\textbf{The ground state of the hydrogen atom.} A basis set of eigenfunctions for the pilot-wave is obtained by solving the equation
\begin{eqnarray} \left(-\frac{1}{2m}\nabla^2+\frac{-e^2}{4\pi\epsilon_0r}\right)\psi(r) = E\psi(r) \end{eqnarray}
where $r$ is the radial distance from the central charge, $m$ the reduced mass of the electron and $E$ is the energy. The ground state solution to this equation is
\begin{eqnarray} \psi_{100}(r) = \left(\frac{1}{\pi a_0^3}\right)^{1/2}e^{-r/a_0} \end{eqnarray}
where $a_0 = \frac{4\pi \epsilon_0}{me^2}$ is the Bohr radius. Hence
\begin{eqnarray} m\dot{\vec{x}}=\nabla S = 0 \end{eqnarray}
and so the electron is at rest. Note that this is not in conflict with results from SQM as far as any measurement of the momentum is concerned. A measurement is a dynamical process which may affect the velocity of the particle. More crucially however, it is a process in which the state of the system affects the velocity components corresponding to the state of the measurement apparatus (such as the position of a pointer or dial). While the electron \emph{is} at rest, it will never be \emph{found} at rest. SQM abstains from any result concerning what the momentum of the electron actually \emph{is}. Details regarding the theory of measurement in PWT are found in section \ref{Measurement}.\\

	\subsection{The ensemble distribution and probability} \label{ensemble}
	
	%\subsubsection{The idea of the ensemble}
	
	PWT is an entirely deterministic theory in which particle positions and properties are at all times well defined. SQM on the other hand is manifestly probabilistic as well as empirically successful. The probabilistic nature of SQM arises in the question where we (the experimenter) will \emph{find} a particular particle, say, or what we \emph{find} some particular value associated with a property of the particle to be. To ask what the position of a particle \emph{is}, is meaningless.
	
	This difference between PWT and SQM regarding the question where particles \emph{are} independently of where they are found, is not problematic as it is \emph{ex hypothesi} not observable and so we cannot use it to empirically distinguish between the theories. The question that remains however is how, if PWT is true, it can account for the success of the manifestly probabilistic theory of SQM in answering questions about where particular particles are \emph{found}.
	 
	Consider what exactly it is that needs to be explained. According to SQM, if a state $\ket{\psi}$ is a non-trivial superposition of eigenstates of a Hermitian operator $\hat{A}$ corresponding to some ``observable'' $A$, i.e.\ if
\begin{eqnarray} \ket{\psi} = \sum_{k=1}^n c_k\ket{\phi_k}	\end{eqnarray}
where $c_k$ is the coefficient of the $k$th eigenstate $\ket{\phi_k}$ of $\hat{A}$ and $\ket{\psi}$ is normalised ($\sum_{k=1}^n|c_k|^2=1$), and if $A$ is ``measured'', then the outcome of the measurement is the $j$th eigenvalue $a_j$ of $\hat{A}$ with probability $|c_j|^2$. This is known as the Born rule. Experimentally however it is impossible to determine the probability of any particular outcome by observing a single system. If we have a system in state $\ket{\psi}$, perform a measurement of $A$ (whatever this requires us to do physically) and obtain outcome $a_j$, then we still know hardly anything at all about the probability distribution of the possible outcomes of the experiment except that $c_j\neq 0$. The way the Born rule is experimentally corroborated is by performing the measurement on a large number of similar systems in state $\ket{\psi}$. That is, by considering an \emph{ensemble} of physical particles (or whatever our system consists of). What we observe is never probability directly but only \emph{relative frequencies}. 

	Consider as an example a two-slit experiment where photons of wavelength $\lambda$ are emitted from a source towards a screen (which is watched by some human observer) and a barrier with two narrow slits (with width $d\ll\lambda$) is positioned between the source and the screen. We observe a diffraction pattern formed by the flashes on the screen and we observe this pattern even when the source is set to only emit individual photons temporally well separated from one another. The way SQM accounts for this result is in terms of a wave function $\psi(\vec{x},t)=\inpr{\vec{x}}{\psi(t)}$ that obeys the Schr\"odinger equation (and hence depends on \emph{both} slits) and evolves such that the relative frequency of flashes at position $\vec{z}$ on the screen is proportional to $|\psi(\vec{z},t_{flash})|^2$. Thus, by the Born rule, the probability of \emph{any individual photon} to be observed at $\vec{z}$ is $|\psi(\vec{z},t_{flash})|^2$ and hence for a large ensemble of photons the relative frequency of flashes at $\vec{z}$ approaches $|\psi(\vec{z},t_{flash})|^2$ and so the diffraction pattern is recovered.
	
	Objective probability in SQM implies that the correct frequencies of experimental outcomes are observed. However, the reverse is obviously not true. A statistical spread of measurement outcomes does not imply the existence of objective probability. For example, these frequencies also follow from the right statistical spread in the initial state across the ensemble of systems together with the right kind of laws of evolution. The fact that we cannot predict with certainty where a particular photon will hit the screen is only due to our ignorance of the initial state of the individual system. This is the approach taken by PWT. 
	
	Before commencing with a more formal presentation of the ensemble distribution $\rho$ in PWT, note that some accounts of the theory present $\rho$ as the distribution of a ficiticious ensemble that expresses our ignorance of the system's particle position, i.e.\ our ignorance of the exact state of the system prior to any measurement procedure. This however seems ill-motivated. The only thing that can possibly justify any particular ficticious distribution is the observed frequency of outcomes of experiments performed on similar systems. The apparent advantage of the ficticious ensemble approach is that for unique systems we can more meaningfully talk of the subjective probability of outcomes for that system (which is likely to sound reassuring to disciples of SQM). However, we have no justification to choose any particular distribution for our ficticious ensemble. In fact, we do not have to either, since the uniqueness of the systems implies that there never will be enough -- indeed any -- data to calculate (subjective or objective) probabilities with any substantial confidence. Jumping ahead somewhat, note also that for quantum non-equilibrium (i.e.\ distributions $\rho\neq|\psi|^2$) to be physical meaningful at all, a real and not ficticious ensemble is required.
	
	%\subsubsection{Evolution of the ensemble distribution}
	
	Suppose then that we have an ensemble of similar systems (that is, systems with the same pilot-wave function) with a density distribution $\rho(\vec{x},t)$ in configuration space. We assume that the ensemble is large enough that $\rho(\vec{x},t)$ can be approximated as a smooth function. Since the guidance equation can be understood as defining a velocity field $v^{\psi}$ for any particular pilot-wave $\psi(\vec{x},t)$, the density function $\rho(\vec{x},t)$ obeys the continuity equation
	\begin{eqnarray} \frac{\partial \rho(\vec{x},t)}{\partial t}+ \nabla\cdot(\rho\dot{\vec{x}}) = 0 ,\end{eqnarray}
where $\nabla$ without a subscript is the $3N$-dimensional derivative operator.
	We may also rewrite equation \ref{ImaginarySE} as
\begin{eqnarray} \frac{\partial |\psi(\vec{x},t)|^2}{\partial t}+\nabla\cdot(|\psi(\vec{x},t)|^2\dot{\vec{x}}) = 0. \end{eqnarray}
Hence $\rho(\vec{x},t)$ and $|\psi(\vec{x},t)|^2$ obey the same first-order partial differential equation. The following theorem follows immediately:
\newtheorem*{Theorem}{Theorem}
\begin{Theorem} If $\rho(\vec{x},t_0)=|\psi(\vec{x},t_0)|^2$ for some time $t_0$, then $\rho(\vec{x},t)=|\psi(\vec{x},t)|^2$ for all times $t$. \end{Theorem}
In words, if the ensemble distribution density and the pilot-wave function amplitude squared satisfy the same boundary conditions, they will be equal at all times. As we will illustrate in our discussion of measurement in the next section, this is sufficient to recover the empirical equivalence of SQM and PWT.

	%\subsubsection{Quantum (non-)equilibrium}

$\rho(\vec{x},t)=|\psi(\vec{x},t)|^2$ is required to reproduce the predictions of SQM. Some have suggested that we must therefore include this constraint as a postulate in the theory. Having a law-like boundary condition seems peculiar and has caused some controversy. However, Valentini\footnote{1991a,b, \cite{ValentiniHTheorem}} has shown that for $\rho(\vec{x},t)\neq|\psi(\vec{x},t)|^2$ a theorem analogous to Boltzmann's H-theorem can be derived, proving that $\rho(\vec{x},t)\rightarrow|\psi(\vec{x},t)|^2$ as $t\rightarrow\infty$. In analogy with thermal equilibrium, the term ``quantum equilibrium'' is therefore appropriate to describe the distribution $\rho(\vec{x},t)=|\psi(\vec{x},t)|^2$. The quantum H-theorem implies that we do not have to impose any special boundary condition but can arrive at the result $\rho(\vec{x},t)\approx|\psi(\vec{x},t)|^2$ by a typicality argument. Valentini has also suggested that quantum non-equilibrium might be found in remnant particles that decoupled in the very early universe (before they had enough time to reach quantum equilibrium).\footnote{e.g. 2007, \cite{ValentiniAstroTests}} Some quantum non-equilbrium distributions would allow phenomena such as superluminal signalling, the distinction of non-orthogonal quantum states and undetected eavesdropping on quantum key distribution.\footnote{Valentini 2002, \cite{ValentiniSubquantumMeasurement}} In general, quantum non-equilibrium would be experimentally distinguishable from equilibrium. However, for the rest of this work we will assume that we are in (or practically indetectably close to) quantum equilibrium, if not indicated otherwise.

	In our discussion so far we have not yet given a dynamical description of how outcomes are obtained, that is, a description of measurement. This is the topic of the next subsection.\\

	\subsection{Measurement}\label{Measurement}
	
	The term ``measurement'' refers to a particular type of interaction between a target system (the system of which some property is to be determined) and the apparatus whose task is to provide some \emph{output} that is in some way dependent on the target system. This output may have various forms but it is perhaps most convenient to think of it as the position of a pointer or a digital display. Note that the concept of measurement is fundamentally anthropocentric. Without any intelligent agent who is observing and able to interpret the output there is no measurement. Without such agents there are only systems that happen to have pointers and digital displays whose state in some way depends on the state of the target system. Agents themselves are only special kinds of (admittedly rather complicated) dynamical systems too, which may be seen as the origin of the measurement problem in SQM because members of this particular class of systems are involved in dynamics that do not supervene on the dynamics of the constituent parts. In PWT there is no such problem. Measurement is simply a term used to describe a particular type of interaction between systems.
	
	We will describe a simple model of measurement where a system interacts with an apparatus with a pointer whose position may be specified by a single variable $y$. Note that all measurement outputs are expressible in terms of position (sound waves are a phenomenon emergent from particle positions, for example). We could also include the experimentalist and model his or her experience of the observation of the output by a variable describing the configuration in his or her brain.
	
	Suppose then that the degrees of freedom of the target system can be uniquely summarised by an $n$-dimensional vector variable $\vec{x}$. At time $t=0$ the pilot-wave function of the system is $\psi_0(x)$. Our measurement apparatus has a pointer moving in one dimension and its position is specified by $y$ (although generalisations to more dimensions are easily obtainable) and initially the pilot-wave function associated with this pointer is a narrow wavepacket $g_0(y)$. Suppose further we wish to perform a measurement of what in the context of SQM would be called an observable associated with the Hermitian operator $\hat{Q}$. We now show that the apparatus is the right tool for a measurement of that ``observable'' if the interaction Hamiltonian between the system and the apparatus is
	\begin{eqnarray} \hat{H}_{int}=a\hat{Q}\hat{p}_y \end{eqnarray}
where $\hat{p}_y=-i\frac{\partial}{\partial y}$ is the momentum operator conjugate to $y$ and $a$ an appropriate constant. Note that in reality this Hamiltonian encodes a very complicated dynamics between the atoms of the system and those making up the apparatus.

	Denote the pilot-wave of the joint system by $\pi(x,y,t)$, with $\pi(x,y,0)=\psi_0(x)g_0(y)$. We make the further assumption that the systems interact for only a very short period of time $\delta t$ (an ``impulse measurement'' in Bohm's terms), so that we can neglect the free evolution of the system and the apparatus. We can do so by scaling the interaction strength $a$ accordingly. The Schr\"odinger equation for the system is then given by
	\begin{eqnarray} i\frac{\partial}{\partial t}\pi(x,y,t) = \hat{H}_{int}\pi(x,y,t), \end{eqnarray}
which is solved by
	\begin{eqnarray} \pi(x,y,t) = e^{-i\hat{H}_{int}t}\pi(x,y,t) = e^{-i\hat{H}_{int}t} \psi_0(x)g_0(y). \end{eqnarray}

	For small time periods $\delta t$ this becomes
\begin{eqnarray}\notag \pi(x,y,\delta t)
&=& (1-i\hat{H}_{int}\delta t)\psi_0(x)g_0(y) \\
&=& \notag \left(1-a\delta t \hat{Q}\frac{\partial}{\partial y}\right)\psi_0(x)g_0(y)\\
&=& \psi_0(x)g_0(y)-a\delta t \hat{Q}\psi_0(x)\frac{\partial}{\partial y}g_0(y).
\end{eqnarray} 
We now write the pilot-wave function of the system as a linear sum of eigenfunctions $\phi_k(x)$ of $\hat{Q}$ with coefficients $c_k$ and eigenvalues $q_k$:
\begin{eqnarray} \psi_0(x) = \sum_k c_k \phi_k(x) \end{eqnarray}
such that
\begin{eqnarray} \hat{Q}\phi_k(x) = q_k\phi_k(x). \end{eqnarray}
However, unlike in SQM, no particular physical meaning is attached to these eigenfunctions. Expressing $\psi_0(x)$ in this form is purely a mathematical convenience. There is no kind of ``eigenvalue realism''(Valentini) or physical ``superposition''. It follows that
\begin{eqnarray} \notag \pi(x,y,\delta t)
&=& \psi_0(x)g_0(y) - \sum_k a \delta t c_k q_k \phi_k(x)\frac{\partial}{\partial y}g_0(y) \\
&=& \sum_k c_k \phi_k(x) \left[g_0(y)-a \delta t q_k\frac{\partial}{\partial y}g_0(y)\right]. \end{eqnarray}
We note that the term in square brackets is of the form $g_0(y)+\delta y\frac{\partial g_0(y)}{\partial y}$ where $\delta y = -a\delta t q_k$ is small since $\delta t$ is small. Hence to a very good approximation
\begin{eqnarray} g_0(y)+\delta y\frac{\partial g_0(y)}{\partial y} = g_0(y+\delta y) = g_0(y-a\delta t q_k) \end{eqnarray}
and so
\begin{eqnarray} \pi(x,y,\delta t) = \sum_k c_k \phi_k(x)g_0(y-a\delta t q_k). \end{eqnarray}

	We see that at the end of the interaction period the pilot-wave function can be expressed as a linear sum of terms such that for the $k$th term the wavepacket of the pointer has been shifted by a distance $a\delta t q_k$. For strong enough interactions ($a$ large enough) this means that the wave packet descriptions contained in the individual terms will not overlap for non-degenerate $q_k$. This implies that the $n+1$-dimensional pilot-wave function of the combined system is non-zero only inside a set of disjoint areas $A_k$ of configuration space.
	
	Recall now that in quantum equilibrium the ensemble distribution density for this combined system is $\rho(\vec{x},y,t)=|\pi(\vec{x},y,t)|^2$. Hence the probability that the particle of an arbitrarily selected individual system $\mathfrak{S}$ of the ensemble is in a particular configuration space volume $\vec{\delta x}\otimes\delta y$ at time $t$ is 
	\begin{eqnarray}\notag \text{Prob}\left(\vec{x}_{\mathfrak{S}}\in(\vec{x},\vec{x}+\vec{\delta x}),y_{\mathfrak{S}}\in(y,y+\delta y);t\right) 
			 &=& \int_{\vec{x},y}^{\vec{x}+\vec{\delta x},y+\delta y} dyd^nx\text{ } \rho(\vec{x},y,t) \\ 
\notag &=& \int_{\vec{x},y}^{\vec{x}+\vec{dx},y+\delta y} dyd^nx\text{ } |\pi(\vec{x},y,t)|^2.\\ \end{eqnarray}
It follows then that the probability that the particle is inside $A_k$, that is, that the pointer is pointing to the $q_k$-outcome, is
	\begin{eqnarray} \text{Prob}\left((\vec{x},y)\in A_k;t\right) = \int_{A_k} dyd^n x\text{ }|\pi(\vec{x},y,t)|^2 = |c_k|^2,  \end{eqnarray}
if $\psi_0$ and $g_0$ are correctly normalised. This is exactly the probability (and hence relative frequency) that we would have derived using the Born rule in SQM. Observation of the pointer changes our knowledge of the position of the particle and so after the measurement we would only consider the reduced ensemble whose member particles agree with the measurement outcome. This constitutes an effective collapse of the wave function. We have thus illustrated that for quantum equilibrium the predictions of SQM are recovered.

	With this derivation we have not only shown the empirical equivalence of SQM and equilibrium PWT, but have also shown that the interaction with the measurement apparatus is described by an effective Hamiltonian $\hat{H}_{int}=a\hat{Q}\hat{p}_y$. In this sense we have provided a dynamical answer to the question of what it means to measure the ``observable'' associated with $\hat{Q}$. Contrast this to SQM where measurement is understood as a dynamically fundamental process.\\

	\subsection{Spin}\label{Spin}
	
	A treatment of spin in PWT is surprisingly simple. Note that, as we shall see, spin in PWT is a property of the pilot-wave, not the particle (the configuration). The space on which the pilot-wave is defined is unchanged if spinless physical particles are replaced by physical particles with spin. This might at first be counterintuitive, especially if we take the term ``particle'' too literally, as we generally speak of particles possessing a certain spin. Note further that this also implies that there is no ensemble of different ``spin configurations'' to account for the subjective uncertainty required to recover the phenomenology of SQM. All uncertainty continues to originate in the ensemble distribution across configuration space, that is, in our uncertainty of the initial spatial position of the physical particles. Recall that all measurement is in some sense a measurement of position. Spin, being a property of the guiding wave, may be observed through its effect on the particle trajectories, for example in the form of a Stern-Gerlach device.
	
	Before presenting a full treatment of spin, we will look at a toy model description of a spin measurement process first found in Bell (1987).\footnote{in \emph{Quantum mechanics for cosmologists}, \cite{Bell1987}, p. 127ff} The simplification here is that we ignore the spatial degrees of freedom of the physicle particle entirely and give a description of spin measurement with a single-variable pointer and a suitable interaction Hamiltonian. Furthermore, we limit ourselves to spin-$\frac{1}{2}$ particles.
	
	The pilot-wave of a spin-$\frac{1}{2}$ particle is a two-component wave function. The only spatial degree of freedom is a single variable $y$ denoting the position of a pointer on our measurement apparatus. Initially the pointer and the spin are independent. The initial wave function $\psi$ is therefore
	\begin{eqnarray} \psi_m(t_0) = \phi(y)a_m \end{eqnarray}
where $\phi(y)$ denotes the initial wave function of the apparatus and we take this to be a narrow wave packet centred at a neutral pre-measurement position $y=0$. $a_m$ is the ``wave function'' in zero dimensions (i.e.\ $a_m\in\mathds{C}$) of the measured particle and $m$ takes the values $\pm1$. The interaction Hamiltonian is given by
	\begin{eqnarray} \hat{H}_{Bell} = -ig(t)\sigma \frac{\partial}{\partial y} \end{eqnarray}
where $\sigma$ denotes the Pauli matrix associated with the appropriate spin measurement and $g(t)$ is a time-dependent coupling constant. For simplicity, take $g(t)=0$ for $t<t_0$ where $t_0$ is the point in time when measurement commences, and $g(t)=g$ is constant thereafter. The Schr\"odinger equation for a spin measurement in the $z$-axis is given by
	\begin{eqnarray} \frac{\partial}{\partial t}\psi_m = -g(t)Z_{mn}\frac{\partial}{\partial y}\psi_n \end{eqnarray}
with $Z=\twomatrix{1}{0}{0}{-1}$ denoting the diagonal Pauli matrix. The sum over the repeated spin indices is assumed. Its general solution for $t>t_0$ in component form is given by
	\begin{eqnarray} \psi_{\pm}(y,t)= e^{-i\hat{H}_{Bell\pm}(t-t_0)}\psi_{\pm}(y,t_0) 
	=e^{\mp g \frac{\partial}{\partial y}(t-t_0)}\psi_{\pm}(y,t_0). \end{eqnarray}
Infinitesimally the evolution at $t_0$ is therefore
	\begin{eqnarray} \psi_{\pm}(y,t_0+\delta t)
	&=& \notag (1\mp g\delta t \frac{\partial}{\partial y})\psi_{\pm}(y,t_0) \\
	&=& \psi_{\pm}(y,t_0)\mp g \delta t \frac{\partial}{\partial y}\psi_{\pm}(y,t_0) \end{eqnarray}
which once more can be understood as an expansion around $t_0$ to first order in $\mp g\delta t$, analogous to our previous discussion of measurement in section \ref{Measurement}. Hence we obtain
	\begin{eqnarray} \psi_{\pm}(y,t_0+\Delta t) = \psi_{\pm}(y\mp g\Delta t, t_0) = \phi(y\mp g\Delta t)a_{\pm} \end{eqnarray}
where $\Delta t$ is finite and positive. Thus the two component wave packets of the pointer associated with corresponding components of the particle wave function $a_m$ separate. 
 
	For the spin-$\frac{1}{2}$ case, the quantum equilibrium distribution corresponds to
	\begin{eqnarray}\label{Bellspinequilibrium}\notag \rho(y,t) &=& |\psi(y,t)|^2 = \psi^*_m(y,t)\psi_m(y,t)\\
	&=&|a_+|^2|\phi(y-g\Delta t)|^2+|a_-|^2|\phi(y+g\Delta t)|^2 \end{eqnarray}
 for $t>t_0$ and $\Delta t=t-t_0$ as before. Again (and henceforth always) the sum over spinor indices is implicit. The results from section \ref{ensemble} apply, as can easily be verified. The current defined in equation \ref{current} is derived from 
	\begin{eqnarray} \frac{\partial}{\partial t}|\psi(y,t)|^2 
	&=& \notag \frac{\partial}{\partial t}(\psi^*_m(y,t)\psi_m) \\
	&=& \notag \left(\frac{\partial}{\partial t}\psi^*_m(y,t)\right)\psi_m(y,t)+ \psi^*_m(y,t)\left(\frac{\partial}{\partial t}\psi_m(y,t)\right) \\
	&=& \notag -g\left[\frac{\partial}{\partial y}(\psi^*_+(y,t)\psi_+(y,t))-\frac{\partial}{\partial y}(\psi^*_-(y,t)\psi_-(y,t)\right] \\
	&=& -\frac{\partial}{\partial y}[\psi^*_m(y,t) g Z_{mn}\psi_n(y,t)],	\end{eqnarray}
where in the third step the Schr\"odinger equation has been used.	The current is therefore
	\begin{eqnarray} j(y,t) = \psi^*_m(y,t) g Z_{mn}\psi_n(y,t). 	\end{eqnarray}
	
	According to equation \ref{guidanceequation} the motion of the configuration space particle (and hence the pointer)\footnote{Strictly speaking we should distinguish between the coordinates in configuration space and those of the pointer in 3-space. However, since we are considering a system with only one effective degree of freedom, using the same coordinate symbol $y$ is unproblematic and avoids unnecessary complexity through the introduction of further symbols.} is then given by
	\begin{eqnarray}\notag \frac{dy}{dt} &=& \frac{j(y,t)}{\rho(y,t)} \\
	\notag &=& \frac{\psi^*_m(y,t) g Z_{mn}\psi_n(y,t)}{\psi^*_m(y,t)\psi_m(y,t)}\\
	\notag &=& g \frac{\psi^*_+(y,t)\psi_+(y,t)-\psi^*_-(y,t)\psi_-(y,t)}{\psi^*_+(y,t)\psi_+(y,t)+\psi^*_-(y,t)\psi_-(y,t)}\\
	&=& g\frac{|a_+|^2|\phi(y-g\Delta t)|^2-|a_-|^2|\phi(y+g\Delta t)|^2}{|a_+|^2|\phi(y-g\Delta t)|^2+|a_-|^2|\phi(y+g\Delta t)|^2}.		\end{eqnarray}

	We can now see how this reproduces the experimental predictions for a spin measurement in SQM. Once the $\phi$-wave packets have separated, the guidance equation reduces to 
	\begin{eqnarray}\label{BellToyEOM}\frac{dy}{dt}=\pm g,\end{eqnarray} 
giving two possible trajectories (the pointer will move either in the positive or negative $y$-direction). The configuration particle will have travelled a distance $\Delta y = \pm g \Delta t$, which is exactly the velocity of the corresponding wave packet. Hence the particle will travel at a constant speed as its position remains unchanged relative to the wave packet, at least for $\Delta t$ sufficiently short that the free evolution (which causes the wave packet to become less localised) can be ignored. 
	
	Equation \ref{Bellspinequilibrium} is an expression for the ensemble density given by two normalised terms $|\phi(y\mp g\Delta t)|^2$ with coefficients given by $|a_{\pm}|^2$ respectively. Hence, in quantum equilibrium, these coefficients also correspond to the subjective probability of the physical pointer to move in the $\pm$-direction. That is, $|a_{\pm}|^2$ correspond to the probabilities of measuring ``spin up'' and ``spin down'', just as predicted by the Born rule in SQM. This completes our discussion of Bell's toy model for spin. We will make use of it in section \ref{QCinPWT}.
	
	For completeness, we will now provide the tools for a more general and physically accurate treatment of spin, although we will again limit detailed discussion to a single spin-$\frac{1}{2}$ particle and we will not discuss any applications or examples. The general Hamiltonian for a particle of spin $s$ is given by\footnote{see Struyve 2004, \cite{StruyvePhD}, p. 13ff, or Landau \& Lifshitz 1977 \cite{LandauLifshitz1977}}
	\begin{eqnarray} H_{spin} = -\frac{D^2}{2m}-\frac{e\gamma}{2m}\vec{S}^{(s)}\cdot\vec{B}+eA_0+V \end{eqnarray} 
in natural units. Here $e$ is the charge of the particle, $\gamma$ the gyromagnetic factor, $S^{(s)}_i$ are the generators of the $2s+1$-dimensional representation of the rotation group $SO(3)$ satisfying $[S^{(s)}_i,S^{(s)}_j]=i\epsilon_{ijk}S^{(s)}_k$, $\vec{B}=\nabla\times\vec{A}$ is the magnetic field, its scalar product with $\vec{S}^{(s)}$ is understood as a sum over the spatial indices $i=1,2,3$, $A_i$ and $A_0$ together are the electromagnetic 4-potential and $V$ is the sum of all other scalar potentials acting on the particle. $D_i=\frac{\partial}{\partial x_i}-ieA_i$ is the corresponding covariant derivative with Hermitian conjugate $D_i^{\dagger}=\frac{\partial}{\partial x_i}+ieA_i$.

	We will now derive the current $\vec{j}(\vec{x},t)$ for the case $s=\frac{1}{2}$. For clarity of notation, we will leave the arguments $(\vec{x},t)$ of the wave function components $\psi_m$ implicit. The Pauli equation and its Hermitian conjugate are then 
	\begin{eqnarray}\notag i\frac{\partial}{\partial t}\psi_m &=& -\frac{D^2}{2m}\psi_m-\frac{e\gamma}{4m}\vec{B}\cdot \vec{\sigma}_{mn} \psi_n+(eA_0+V)\psi_m\\
	 -i\frac{\partial}{\partial t}\psi^*_m &=& -\frac{D^{\dagger\text{ }2}}{2m}\psi^*_m-\frac{e\gamma}{4m}\vec{B}\cdot \psi^*_n\vec{\sigma}_{nm} +(eA_0+V)\psi^*_m, \end{eqnarray}
choosing $\vec{S}^{(s)}=\frac{1}{2}\vec{\sigma}$ and recalling that the Pauli matrices $\sigma_i$ are Hermitian. Hence
	\begin{eqnarray} \frac{\partial}{\partial t}|\psi|^2
	\notag &=& \left(\frac{\partial}{\partial t}\psi^*_m\right)\psi_m +
						 \psi^*_m\left(\frac{\partial}{\partial t}\psi_m\right) \\
				 &=& \frac{i}{2m}[\psi^*_mD^2\psi_m-\psi_mD^{\dagger 2}\psi^*_m]
				   		+ \frac{ie\gamma}{4m} \vec{B}\cdot \vec{\sigma}_{mn}[\psi^*_m\psi_n - \psi^*_n\psi_m].
	\end{eqnarray}
For the spin case, the current consists of two parts, $\vec{j}(\vec{x},t) = \vec{j}_c(\vec{x},t)+\vec{j}_s(\vec{x},t)$, such that $\vec{j}_c$ resembles the standard Schr\"odinger current
	\begin{eqnarray} 
 	\nabla\cdot\vec{j}_c(\vec{x},t) &=& -\frac{i}{2m}[\psi^*_mD^2\psi_m-\psi_mD^{\dagger 2}\psi^*_m]
	\end{eqnarray}
and $\vec{j}_s$ is an additional spin term that resembles the magnetization current of a classical polarized medium.\footnote{See Struyve (2004), p. 14, \cite{StruyvePhD}} The addition of the spin term has been proposed in work by Holland\footnote{1999, \cite{Holland1999}} and Holland and Philippidis.\footnote{2003, \cite{HollandPhilippidis2003}} We will not provide a detailed treatment here and the reader is referred to the work of these authors.

We will use the common shorthand notation $\partial_i=\frac{\partial}{\partial x_i}$. In the following a sum over repeated spatial indices as well as spinor indices is assumed. For $\vec{j}_c(\vec{x},t)$ we have
	\begin{eqnarray} \nabla\cdot\vec{j}_c(\vec{x},t)
	\notag &=& -\frac{i}{2m}[\psi^*_m(\partial_i-ieA_i)(\partial_i-ieA_i)\psi_m-
							\psi_m(\partial_i+ieA_i)(\partial_i+ieA_i)\psi^*_m] \\
	\notag &=& -\frac{i}{2m}[\psi^*_m\partial_i^2\psi_m-\psi_m\partial_i^2\psi^*_m\\
	\notag		&&\qquad\qquad -2ieA_i(\psi^*_m\partial_i\psi_m+\psi_m\partial_i\psi^*_m)-2ie(\partial_iA_i)\psi^*_m\psi_m]\\
				 &=& -\frac{i}{2m}\partial_i[\psi^*_m\partial_i\psi_m-\psi_m\partial_i\psi^*_m]
				 			-\frac{e}{m}\partial_i[A_i\psi^*_m\psi_m]
	\end{eqnarray}
and so the current is
	\begin{eqnarray} \vec{j}_c(\vec{x},t) = -\frac{i}{2m}(\psi^*_m\nabla\psi_m-\psi_m\nabla\psi^*_m)
																							-\frac{e}{m}\vec{A}\psi^*_m\psi_m,
	\end{eqnarray}
which we may choose to rewrite in index-free notation as
	\begin{eqnarray} \vec{j}_c(\vec{x},t) = -\frac{i}{2m}(\psi^{\dagger}\nabla\psi-(\nabla\psi)^{\dagger}\psi)
																							-\frac{e}{m}\vec{A}\psi^{\dagger}\psi.
	\end{eqnarray}

	For $\vec{j}_s(\vec{x},t)$ we have\footnote{See e.g. Holland (1999), \cite{Holland1999}}
	\begin{eqnarray} \vec{j}_s(\vec{x},t) = \frac{\gamma}{4m}\nabla\times(\psi^{\dagger}\vec{\sigma}\psi). \end{eqnarray}
	With the definition of a spin 3-vector
	\begin{eqnarray} \vec{s}=\frac{\psi^{\dagger}\vec{\sigma}\psi}{\psi^{\dagger}\psi} \end{eqnarray}
	this may be rewritten as
	\begin{eqnarray} \vec{j}_s(\vec{x},t) = \frac{\gamma}{4m}\nabla\times(\psi^{\dagger}\psi \vec{s}). \end{eqnarray}
	
	The law of motion of the configuration space particle is therefore
	\begin{eqnarray} \frac{d\vec{x}}{dt} 
	\notag	&=& \frac{\vec{j}_c+\vec{j}_s}{\psi^{\dagger}\psi} \\
	\notag	&=& \frac{1}{\psi^{\dagger}\psi} \left[ -\frac{i}{2m}(\psi^{\dagger}\nabla\psi-(\nabla\psi)^{\dagger}\psi)
							-\frac{e}{m}\vec{A}\psi^{\dagger}\psi + \frac{\gamma}{4m}\nabla\times(\psi^{\dagger}\psi \vec{s}) \right]\\
	\notag	&=& -\frac{i}{2m\psi^{\dagger}\psi}(\psi^{\dagger}\nabla\psi-(\nabla\psi^{\dagger})\psi)-\frac{e}{m}A +
							 \frac{\gamma}{4m\psi^{\dagger}}\nabla\times(\psi^{\dagger}\psi \vec{s}).\\
	\end{eqnarray}

	\pagebreak
	\subsection{Objections and open questions}\label{objections}
	
	PWT has never been popular among physicists. Some reasons for this may be found in history, such as Pauli's objection at the Solvay Conference (an objection to which PWT can respond satisfactorily), Bohm's association with communism during the McCarthy era and the often dogmatic defense of Copenhagen quantum mechanics by Bohr and his followers, that even made it into prominent textbooks:
	\begin{quote} ``It is clear that this result [the diffraction pattern of a two-slit experiment] can in no way be reconciled with the idea that electrons move in paths.'' (Landau \& Lifshitz 1977, \cite{LandauLifshitz1977}, p. 2) \end{quote}
When PWT is discussed in the literature, one finds various objections that supposedly illustrate fallacies in the theory. While there are problems or at least open questions in PWT, good criticism is often hard to find among the numerous poorly constructed arguments that stem from a misunderstanding of the theory or assumptions made in Copenhagen that simply are not part of pilot-wave theory. To list them all is hardly feasible and the purpose of this work is not to promote PWT as \emph{the} superior quantum theory, or any such agenda (although I do hope to contribute to a clarification of some of the many issues and myths that have pervaded discussions of PWT). For possible responses to some of the most common criticisms, see Mike Towler's lecture notes on PWT\footnote{\cite{TowlerLectures}} and Passon's 2005 paper\footnote{2005, \cite{Passon2005}}. Some popular objections together with a brief summary for a possible response are listed in the appendix. Here I will treat two points in particular: the (non-)uniqueness of the guidance equation and the question of how 3-space is emergent from the configuration. The ``Everett in denial'' objection defended by Deutsch, Zeh, Brown and Wallace will be treated separately in section \ref{Denial} as there is much to be learned from it, not least for Everettians themselves (independently from Valentini's counterattack based on eigenvalue realism\footnote{2010, \cite{ValentiniInDenial}}).

	The first issue we will discuss here is the apparent ``gauge'' freedom in the current $\vec{j}(\vec{x},t)$. Recall that the current is defined implicitly via the equation
	\begin{eqnarray}\label{currentdef} \frac{\partial}{\partial t}|\psi(\vec{x},t)|^2 + \nabla\cdot\vec{j}(\vec{x},t) = 0. \end{eqnarray}
The following criticism has been made repeatedly: \textbf{Defined in this way, the current $\vec{j}(\vec{x},t)$ may be modified by adding a divergence-free term $\vec{a}(\vec{x},t)$ (with $\nabla\cdot\vec{a}(\vec{x},t)=0$) and still satisfies the defining equation \ref{currentdef}. Hence the path traced out in configuration space is not unique (and so neither are the paths of physical particles in 3-space). This makes it unlikely that these paths are physical.}

	A point of clarification: It can be shown that adding a divergence-free term to the current does not change the predictions of experimental measurement outcomes (as it has no effect on the evolution of the equilibrium ensemble density). However, it does change the individual trajectories and hence if these trajectories are to be physical, this arbitrary term should better be fixed uniquely.
	
	One possible response is, of course, to admit that even if there is a single unique physical trajectory (and hence a single correct current), we will never be able to know which one that is. Just because we cannot know it does not mean that it does not exist. However, stronger responses exist. One possibility of fixing $\vec{a}(\vec{x},t)$ is by appealing to certain symmetry and simplicity constraints. This strategy has been pursued by D\"urr, Goldstein and Zangh\`i\footnote{1992, \cite{DuerrGoldsteinZanghi1992}}. While we will omit the details here, this raises questions about how far principles of symmetry and simplicity should be employed to fix physics: Can we assume \emph{a priori} that such metatheoretic principles concerning the nature of reality should hold? The argument appears to be back to front. Should it not be the physics that determines which principles hold, rather than vice versa?
	
	Another interesting and purely physical response to the question of uniqueness has been put forward by Holland\footnote{1999, \cite{Holland1999}} and again defended by Holland and Philippidis.\footnote{2003, \cite{HollandPhilippidis2003}} They remind us that the non-relativistic treatment of our theory should really be understood as the limiting case of a relativistic treatment. The requirement that non-relativistic PWT be embedded in a relativistic theory uniquely fixes the law of motion.
	
	In particular, consider a relativistic spin-$\frac{1}{2}$ particle whose Dirac current is given by
	\begin{eqnarray} j^{\mu} = \bar{\psi}\gamma^{\mu}\psi \qquad \qquad \text{with } \partial_{\mu}j^{\mu} = 0 \end{eqnarray}
where $\bar{\psi} = \psi^{\dagger}\gamma^0$. In this case the law of motion is given by
	\begin{eqnarray} \frac{dx^i}{dt} = \frac{j^i}{j^0} \end{eqnarray}
where $i=1,2,3$ denote the spatial indices of a 4-vector. Adding a divergence-free term to the current corresponds to the operation
	\begin{eqnarray} j^{\mu}\qquad\rightarrow\qquad j'^{\mu} = j^{\mu}+a^{\mu} \qquad\qquad \text{with } \qquad \partial_{\mu}a^{\mu}=0. \end{eqnarray}
In that the trajectories defined by $j'^{\mu}$ produce the same ensemble density, we require $a^0=0$. Lorentz transforming the current to a different frame with spinors $\psi'(x'^i,t')$ gives a value $j'^0=\bar{\psi'}\gamma^0\psi' = \psi'^{\dagger}\psi'$ for the ensemble density. Again we require $a'^0=0$ but the only 4-vector $a^{\mu}$ whose 0-component vanishes in all frames is $a^{\mu}=0$. This fixes the relativistic law of motion. In the non-relativistic limit the Dirac equation reduces to the Pauli equation for a two-component spinor $\phi^A$. Since the Dirac current is unique, so is its non-relativistic limit, i.e.\ there is no freedom to add a divergence-free term.

	Thus we have fixed the law of motion for non-relativistic spin-$\frac{1}{2}$ particles uniquely. A similar treatment for general spin would reduce the equation for spin-$0$ particles to the standard $\frac{d\vec{x}}{dt}=\frac{1}{m}\nabla S$. One may find it odd that we have to appeal to spin in order to uniquely fix the equation of motion for spinless particles. However, such criticism is ill-founded. Spin-$0$ means just that: a particle with total spin 0, not one to which the concept of spin does not apply. To arbitrarily treat spin-$0$ particles completely differently is unwarranted. 
%_____	
	
	We will now turn to a question concerning the ontology of 3-space. We present the following criticism of PWT: \textbf{In PWT configuration space is not merely a mathematical construct, but part of the fundamental ontology. Yet our world as we experience it has only $3$ and not $3N$ spatial dimensions and the structure of the $3N$-dimensional configuration space is insufficient for its emergence.}
	
	The idea here is this: How, we may ask, does 3-space fit into the framework of PWT? Presumably pilot-wave theorists wish to claim that it emerges from (and thus supervenes on) configuration space. In particular, if fundamentally we have a $3N$-dimensional configuration space, then on it supervenes a world with $N$ identical particles whose positions are each specified by a projection of the configuration particle onto a particular 3-dimensional subspace such that any two such subspaces are orthogonal. But this simply does not follow. Configuration space and the configuration particle by themselves do not specify how the $3N$ dimensions are grouped into $N$ sets of three dimensions. For a configuration $(x_1,x_2,x_3,x_4,x_5,x_6)$ in six dimensions, are the two particles at $(x_1,x_2,x_3)$ and $(x_4,x_5,x_6)$ or at, say, $(x_1,x_3,x_5)$ and $(x_2,x_4,x_6)$? Or why should the first particle be at $(x_1,x_2,x_3)$ and not $(x_2,x_3,x_1)$ or some other permutation thereof? And finally, why would a world emergent from a $3N$-dimensional configuration space be one of $N$ particles in $3$ dimensions and not $3$ particles in $N$ dimensions or even $3N$ particles in one dimension?%\footnote{A further question is: How can the configuration space/configuration particle account for different particle masses? However, if we are talking about the emergence of particle mass, perhaps we ought to ask this question in the context of pilot-wave field theory. I am not claiming that a field-theoretic treatment solves the problem of mass, but it seems that in order to be able to respond to the problem correctly, results concerning mass emergence from symmetry breaking etc.\ ought to be taken into account. Since our treatment here is entirely non-relativistic, I will not discuss this question here.}
	
	One possible response would be to show that a natural association of particles to 3-positions follows from the type of potentials acting between them. In particular, potentials with terms of the form $V = a|\vec{x}_1-\vec{x}_2|^b$ clearly ``identify'' particles: If $\vec{x}_1 = (x_1,y_1,z_1)$ and $\vec{x}_2 = (x_2,y_2,z_2)$ each specify the position of a particular particle, then the potential term is proportional to the \emph{distance} between particles to some power. If instead we were to think of three particles in two dimensions, say, then the potential, while a perfectly good mathematical expression, loses this intuitive meaning and may even become dependent on the choice of coordinate system. However, this does not fully alleviate our concerns since we may still wonder what happens in the absence of such potentials and also whether perhaps other ``natural'' interpretations are possible for certain associations of physical particle coordinates with configuration space coordinates. Nevertheless an approach using potentials seems promising, although some more work needs to be done in this area. For an account on using potentials for identifying the correct associated 3-space, see Albert's \emph{Elementary Quantum Metaphysics}\footnote{1996, \cite{Albert1996}}.
	
	If no fully satisfactory response to this problem is found, it would follow that PWT simply does not get away with postulating only the pilot-wave and the particle in configuration space, but more structure is required. It may seem an obvious alternative to propose therefore that 3-space itself is also fundamental. However, this would merely change the problem. If both configuration space and 3-space are fundamental, then \emph{a priori} the two are unrelated. For our theory to accurately describe our world, we would further have to postulate a connection between the two, namely that the positions of particles in 3-space corresponds exactly to the configuration in configuration space. Thereby we would automatically postulate which subspaces of configuration space correspond to which particles. Thus we have not actually avoided the additional ontological baggage of having to postulate structure additional to the configuration space and particle.
	
	%At this point I do not see a satisfactory answer to the problem.\footnote{\label{crazyidea}Here is one bold suggestion (which one may or may not choose to take seriously): \emph{All} possible divisions of configuration space into subspaces of equal dimensions have emergent worlds. Some of them have three spatial dimensions. We just happen to be part of one of them and so from our world-internal point of view we naturally ask such questions. We did \emph{not} postulate all these worlds in addition to the already existing ontology (a fallacy analogous to that made by Everettians for some time), but they are all contained in the structure of the fundamental ontology of PWT. The question here is one of the nature of ``emergence'', not of the ontology itself.}
	One way to avoid the issue altogether is to regard the wave function not as real but as law-like or nomological (a view taken by e.g.\ Goldstein\footnote{2010, \cite{Goldstein2010}}) and disregard the reality of configuration space entirely. Instead, we would take particles in 3-space as fundamental and say that the general law of motion of a system is that its particles move \emph{as if} the point representing its configuration in configuration space (here a merely mathematical construct) were guided according to de Broglie's guidance equation by a wave function obeying the Schr\"odinger equation. 
	
	The idea of a fictitious wave function that only serves as a convenient (maybe even unavoidable) construct strikes one as peculiar. Firstly, as Valentini\footnote{2010, \cite{ValentiniInDenial}} points out, $\psi$ contains a lot of independent and contingent structure. In particular, it evolves in time independently from the evolution of physical particles. This is an argument made even by opponents to PWT as a whole, such as Brown and Wallace.\footnote{2007, \cite{BrownWallace2007}} We cannot simply assign the predicate ``is real'' to an entity when it seems convenient. Criteria such as its role in the theory should be used to determine whether or not it is physical. The properties of the pilot-wave very strongly suggest that it should be understood as a real entity.
	 
	An argument in favour of a law-like pilot-wave is found, for example, in D\"urr et al.\footnote{1997, \cite{DuerrGoldsteinZanghi1997}} They argue that the time evolution of the wave function of some system is merely an illusion since the wave function of the universe as a whole is static and unique. Valentini\footnote{2010, \cite{ValentiniInDenial}} replies that results from quantum gravity suggest that solutions of the universal wave function $\Psi$ satisfying the Wheeler-DeWitt equation and other constraints are not at all unique, although he urges caution in the light of the problem of time in quantum gravity.
	
	 Given that a law-like pilot-wave would resolve the ontological problem of 3-space, it might be tempting to side with defenders of the fictitious wave function. This might be deemed more acceptable if we saw PWT only as a temporary theory to be replaced by another, better one (which was de Broglie's view, although he still saw the pilot-wave as real). Perhaps the pilot-wave \emph{is} dispensable in an improved description of the dynamics. Perhaps the motion of particles can be summarised by a set of concise laws that do not involve any ``moves as if'' description and a fictitious wave in configuration space --- we simply have not found the right formalism to do so yet. If so, we should be happy to give up the idea that configuration space and the entities therein are fundamental. We can then understand PWT as an effective theory. However, in the absence of any such superior theory, PWT by itself suggests that the pilot-wave and configuration space ought to be understood as real, even if this means that the question of the ontology of 3-space and possible additional structure on configuration space remains open.
	 
	 Note that there is a similar problem in Everettian Many-Worlds Theory (MWT). In MWT the wave function is an object in Hilbert space and the question that arises is how the dimensions of Hilbert space correspond to the spatial position of particles in 3-space. Appealing to decoherence might help to select the position basis as a preferred basis, but it does not provide a solution to why 3-space rather than 2-space (say) or how a particular dimension in Hilbert space relates to a particular spatial coordinate of a particle in the supposedly emergent 3-space. The question is once again one of the nature of emergence, which we will treat in greater length in the next section in the context of the ``Many Worlds in Denial'' objection. 
	 
	 Instrumentalist quantum mechanics does not have this problem as it is simply abstinent from anything but measurement outcomes (Hilbert spaces just serve as mathematical tools to predict such outcomes), which is itself unsatisfactory for anyone not taking a strictly positivist point of view.
	 
	 The ontological status of the pilot-wave will, of course, be highly relevant when trying to answer the question what exactly it is that allows for the speed-up of quantum computing.\\

	\subsection{Pilot-Wave Theory, Many-Worlds and Many-Worlds in denial}\label{Denial}
		
	Finally we will examine an attack that has become known as the ``many-worlds in denial'' objection to PWT and has been primarily defended by Deutsch\footnote{1996, \cite{Deutsch1996}}, Zeh\footnote{1999, \cite{Zeh1999}} and Brown and Wallace\footnote{2007, \cite{BrownWallace2007}}.	Brown and Wallace write,
	\begin{quote} ``... the corpuscle's role is minimal indeed: it is in danger of being relegated to the role of an epiphenomenal `pointer', irrelevantly picking out one of the many branches defined by decoherence, while the real story --- dynamically and ontologically --- is being told by the unfolding evolution of those branches. The `empty wave packets' in the configuration space which the corpuscles do not point at are none the worse for its absence: they still contain cells, dust motes, cats, people, wars and the like.'' (Brown and Wallace (2007), p. 8-9) 
	\end{quote}
Deutsch puts it more succinctly:
	\begin{quote} ``[P]ilot-wave theories are parallel-universe theories in a state of chronic denial.'' (Deutsch (1996), p. 225)
	\end{quote}
		
	The criticism relies crucially on the reality of the wave function in PWT. Goldstein's et al.\ view of a nomological wave function would therefore be one way to escape the problem. However, for the following discussion let us assume that the theory in question is PWT with a real ontological pilot-wave.
	
	The essence of the attack is, in brief, that the wave function itself contains all the structure required for the emergence of the 3-space reality with which we are familiar. The configuration space particle, whose purpose is to ``pick out'' a single ``real'' history (trajectory), does not, the argument goes, add anything new and is therefore reduntant. On closer analysis we will discover that the debated issue here is ultimately one concerning the question of emergence.
	
	First, however, I will partially defend MWT against a counter-attack launched by Valentini\footnote{2010, \cite{ValentiniInDenial}}. Valentini criticises that the Everettian's belief in the existence of many worlds is based on the misguided idea of eigenvalue realism. The way MWT (and also SQM) is often presented does indeed lead to this conclusion but if we are to weigh up between PWT and MWT, we should do so using their strongest formulations and MWT can be formulated without assuming eigenvalue realism. 
	Valentini is correct in calling out those who ascribe ``parallel universes'' or any such thing to the different terms in a state in a superposition. Suppose, for example, the quantum state of a system in the MWT description is
	\begin{eqnarray} \ket{\psi} = \sum_k c_k \ket{\phi_k} \qquad\qquad\qquad \text{with} \qquad \sum_k |c_k|^2 = 1
	\end{eqnarray}
where for some Hermitian operator $\hat{A}$, 
	\begin{eqnarray} \hat{A}\ket{\phi_k} = a_k\ket{\phi_k}, \end{eqnarray}
i.e.\ $\ket{\phi_k}$ are eigenstates of $\hat{A}$ with respective eigenvalues $a_k$. The Everettian might make two claims here: (1) There is a set $\{U_k\}$ of parallel universes each of which characterised by a different state $\ket{\phi_k}$. (2) When a ``measurement'' corresponding to $\hat{A}$ occurs, that is, when the system interacts with an ``apparatus'' and other environment in some particular way, then the universe (that is, our branch of the ``multiverse'') splits into several new branches.

	Claim (1) is unwarranted: $\sum_k c_k \ket{\phi_k}$ is a mathematical expression for $\ket{\psi}$. There is no reason to assign any ontological status to the terms $\ket{\phi_k}$. It might be tempting to do so in anticipation of the branch splitting resulting from measurement of $\hat{A}$, but the state would be no different if instead we were about to perform some other measurement corresponding to Hermitian operator $\hat{B}$ and the expression in terms of eigenstates of $\hat{A}$ would be just as valid. The terminology of ``superposition'' only serves to confuse things. Nevertheless MWTists do often assign a reality to individual terms in the mathematical expression of a state. For example, David Deutsch's claim that quantum parallelism in quantum computing (see below) is a result of different computations happening in parallel universes crucially relies on eigenvalue realism. Valentini is correct in condemning such a view.
	
	Claim (2) is misleading. Suppose through the interaction between system $\mathfrak{S}$ and apparatus/environment $\mathfrak{A}$ the following evolution occurs:
	\begin{eqnarray} \left(\sum_k c_k\ket{\phi_k}_{\mathfrak{S}}\right)\ket{0}_{\mathfrak{A}} \qquad \rightarrow \qquad \sum_k c_k\ket{\phi_k}_{\mathfrak{S}}\ket{a_k}_{\mathfrak{A}} \end{eqnarray}
where $\ket{0}_{\mathfrak{A}}$ is some ``neutral'' initial state of the apparatus and $\ket{a_k}_{\mathfrak{A}}$ is an apparatus state that would traditionally be associated with ``measurement outcome'' $a_k$. In the final state the macroscopic apparatus and evironment look as if they are in a superposition, corresponding to the splitting of a single world branch (with $\ket{0}_{\mathfrak{A}}$) into many. Note that there is nothing to stop us from changing our basis of the combined system in such a way that $\sum_k c_k\ket{\phi_k}_{\mathfrak{S}}\ket{a_k}_{\mathfrak{A}}$ is one of its basis states and suddenly there is no more apparent superposition. But this only hides what is happening since $\mathfrak{S}$ and $\mathfrak{A}$ have become entangled and are thus no longer separable. Decoherence may pick out a preferred basis but it does not immediately follow that this basis should be given preferred ontological status.

	The point is this: Mathematically we can express the state in whatever way we like, before as well as after the ``measurement''. It has nothing to do with different worlds or parallel universes. In fact, talking of ``many worlds'' only confuses matters as it raises questions such as what these worlds are.\footnote{It also seems that what has become known as the ``preferred basis problem'' in MWT is not actually a problem but a result of eigenvalue realism and the misleading terminology of many worlds.} What is unambiguous is the evolution of the wave function. It appears that the strongest form of a misnamed ``many worlds theory'' is one that completely omits the talk of worlds and only talks about the wave function. It has nothing to do with eigenvalues of operators either. Claim (2) would be better stated as ``The wave function evolves in a particular way such that the state of (microscopic) $\mathfrak{S}$ and that of (macroscopic) $\mathfrak{A}$ become entangled.'' This leaves MWT a neat theory with a very simple ontology.
	
	The only question that involves ``worlds'' is the question of how the appearance of our 3-space world emerges from the universal wave function. This is where decoherence theory might possibly be needed. Also, the question has nothing to do with possible mathematical representations of the wave function. It is purely physical. Whether the wave function alone is sufficient for the emergence of our 3-space world is, it seems, also the main difference between pilot-wave and many-world theorists.
	
	Let us now discuss the ``many-worlds in denial'' attack on PWT. Firstly, removing the configuration space particle does not reduce PWT to Everett (indeed Bell\footnote{1987, in \emph{Quantum mechanics for cosmologists}, \cite{Bell1987}} calls the theory ``Everett (?)''). Everettian MWT is a theory of a wave function in \emph{Hilbert space}, PWT's pilot-wave is an object in \emph{configuration space}. While for some systems Hilbert space in the position basis is isomorphic to configuration space, the difference is evident when considering spin, for example. A spin-$\frac{1}{2}$ particle has a Hilbert space $\mathfrak{H}_{pos}\otimes\mathfrak{H}_{\frac{1}{2}}$ and only the first subspace here corresponds to configuration space. But let us give Deutsch, Zeh, Brown and Wallace the benefit of the doubt. 

	Given MWT's neater ontology (only wave function instead of wave function plus particle), what might possibly induce us to prefer PWT? What comes to mind first and foremost is that, unlike MWT, it gives a conclusive account of how the apparently probabilistic measurement outcomes are recovered.\footnote{In recent years MWTists have distinguished between the ``Incoherence Problem'' (How can we even make sense of probability in MWT?) and the ``Quantitative Problem`` (How do we recover Born rule probabilities?). But recall that what needs to be explained is only the statistical frequency of outcomes. We do not have to make sense of real probability at all, neither in PWT nor MWT. The apparent need for probability is a remnant of ideas of Copenhagen QM. Hence there is no Incoherence Problem. Regarding the Quantitative Problem, Deutsch (1999), \cite{Deutsch1999}, and Wallace (e.g. 2009), \cite{Wallace2009} have suggested a decision theoretic approach. However, even if the approach succeeds, what it does is give an account of how rational agents should act according to credences based on the Born rule; it does not explain why we appear to have records of experiments whose outcomes match the Born rule (in other words, why we are part of a branch of the wave function in which this correspondence between past frequencies and initial states of our experiments holds).} But this is not the heart of the issue. In fact, MWTists will deny that the PWT strategy of explaining probabilities succeeds as it relies on the assumption that the particle really does pick out the ``real'' configuration.\footnote{Note however that if we do observe quantum non-equilibrium in relic particles or elsewhere as suggested by Valentini (2007, \cite{ValentiniAstroTests}), Everettians will have to reconsider their position.} In particular, they claim that since all the structure is in the wave function, those other ``empty waves'' are just as real (and hence the particle and the particle ensemble distribution add nothing at all to the theory).
	
	This raises several issues, in particular regarding the nature of these empty waves. Consider Deutsch's remarkable point that the particle in configuration space
	\begin{quote} ``...\ occupies one of the `grooves' in that immensely complicated multidimensional wave function. The question that pilot-wave theorists must therefore address ... is what are the \emph{unoccupied} grooves?'' (Deutsch 1996, p. 225) 	\end{quote}
	Deutsch suggests that these unoccupied grooves must themselves be real. This is a different point altogether! Here the suggestion is not that the trajectories are not real and the wave function is all there is, but that \emph{all} mathematically possible trajectories are real. While this certainly constitutes a many-worlds theory, it is nothing like Everettian MWT.\footnote{Furthermore, it is not the kind of many-worlds theory to which Deutsch is referring when discussing ``quantum parallelism'' in the context of quantum computations.} The ``grooves'' do not intersect (as the guidance equation is first order), so there is no splitting or merging of branches in any sense at all and it suddenly becomes much more meaningful to talk about ``parallel worlds''. 
	
	All this said, perhaps this theory of many de Broglie-Bohm worlds may even be an acceptable alternative to the PWTist. Note that the trajectories form a dense set in configuration space (the union of all trajectories is of fractal dimension $3N$) and thus are uncountable. \emph{A priori} no density measure is defined on this ensemble. Perhaps we could imagine a many-de Broglie-Bohm-worlds theory with an ensemble density $\rho_{traj}=|\psi|^2$? We previously (section \ref{ensemble}) pointed out that PWT does not require an ensemble of universes because no relative frequencies of outcomes could ever be obtained. However, this does not \emph{rule out} an ensemble of universes and here we have a theory that has just that. 
	
	%Let us therefore digress to consider this idea in some more detail. Firstly, let us show (in a very much simplified way) that $\rho_{traj}=|\psi|^2$ actually accounts for Born rule probabilities for measurements \emph{within} some particular world. suppose that a system $\mathfrak{S}$ characterised by a variable $y$ is initially sufficiently isolated that its interaction with the environment $\mathfrak{E}$, characterised by a variable $r$, is negligible. The wave function is then given by
%	\begin{eqnarray} \psi(r,y,t) = \phi(r,t)\chi(y,t) \end{eqnarray}
%where $\chi$ and $\phi$ are the wave functions of $\mathfrak{S}$ and $\mathfrak{E}$ respectively. CONTINUE HERE
	
  On the one hand such a theory's ontology is in a sense much richer than that of standard PWT: One trajectory (world) is replaced by a continuous infinity thereof. On the other hand, the metaphysical question ``Why this particular trajectory and not some other?'' can now be answered anthropically. But does this enrichment of ontology matter, given that no new \emph{types} of entities are postulated? And what is the status of $\rho_{traj}=|\psi|^2$? Valentini\footnote{2010, \cite{ValentiniInDenial}} considers it natural to reduce the ensemble to a single trajectory (i.e.\ to return to standard PWT). I tend towards agreement with Valentini here, although ultimately various philosophical considerations would come to play. We will leave many-de Broglie-Bohm-world theories as an interesting curiosity.
	
	Brown and Wallace's claim is different from Deutsch's. They argue that since in PWT all the structure is already contained in the wave function we should do away with the configuration space particle.\footnote{It becomes once again obvious in this discussion how misleading the term ``particle'', or even ``corpuscle'', is. We should better talk of a single selected ``configuration'' and its evolution. However, in order not to deviate too far from the literature, I will continue to use the standard terminology.} We should question their premise here. The structure that makes up ``cells, dust motes, cats, people, wars and the like'' is not in the wave function but, if anything, in the configuration space itself. A point in configuration space corresponds to a particular configuration, in which these macroscopic concepts are obtained through some kind of coarse-graining. What the value of the wave function $\psi$ is at that point has nothing to do with it. In particular, given that the amplitude of $\psi$ has apparently nothing to do with any kind of ``degree of reality'' according to these authors, why should a zero amplitude suddenly have a very different ontological meaning? $\psi$ is defined even where $|\psi|=0$.\footnote{Of course, at such points the phase $S$ is not defined, but declaring definability of phase as \emph{the} defining factor of what is real seems peculiar to say the least.}
	
	Perhaps MWTists will respond by saying that the \emph{space} (Hilbert or configuration) is not real but just a mathematical tool to represent the quantum state (the wave) mathematically as a function. How configuration or Hilbert space might contain the structure corresponding to the structure found in 3-space is relatively easy to see, although we have discussed problems associated with this correspondence in the previous section. Yet to explain the emergence of our 3-space world from an entity whose properties we can only express when representing it in some purely mathematical space is not straightforward. PWTists claim that the wave function simply is not enough to account for this emergence, with or without decoherence. They therefore see the need to postulate the existence of a preferred (read, ``realized'') configuration.\footnote{To ask how the corpuscle/particle does this seems out of place and is a question resulting from misinterpretating it as some entity that is anything but a realized configuration.} 
	
	The exposition of these issues here has been somewhat brief and a lot more could be said. However, we conclude the question at the heart of the debate between PWTists and MWTists is one of emergence. MWT is build on a reduced ontology relative to PWT, but has an apparent explanatory gap. If this gap can be filled (for example, by relating wave function amplitudes to the reality of structure in a meaningful way) it is conceivable that this might also provide a neat solution to open question of probability (or rather relative frequency) in MWT. In that case, we should all become Everettians. Meanwhile however we might as well accept a richer ontology in order to bridge this explanatory gap.
	
	This completes our discussion of pilot-wave theory. We will return to it in section \ref{QCinPWT}.

%______________________________________________________________________________________________________________________
		
	\pagebreak		
	\section{Quantum Computing}\label{QC}
			
	\subsection{Physical computations}

	Before we can proceed with an analysis of quantum algorithms in the PWT-picture, some groundwork has to be done. Quantum computing as a research area is now over two decades old and there is a large number of introductory texts of varying length, thoroughness and difficulty. Today quantum computing is a vast field, both theoretical and experimental, and also widely known and appreciated. A detailed review here is therefore neither feasible nor necessary. However, there are some particular points that ought to be emphasised in preparation for the work of the subsequent sections.
	
	Fundamentally, there is the question what we mean by a computation. An attempt of a definition is this: 	
	\newtheorem*{Computation}{Definition}
	\begin{Computation}A computation is the physical evolution of a system according to some dynamics from a known initial condition such that during the process one or more output values are obtained, and there is a known map $\mathfrak{I}$ from these output values to possible solutions to the problem posed.
	\end{Computation}
Nature is full of physical processes. Systems evolve from some initial state to some later state according to some pattern that can be codified in mathematical formalism as a set of equations of motion or more generally as a set of dynamical laws. Yet these processes do not constitute computations, even if the system in question interacts with some form of apparatus which subsequently causes a screen or other output device to display a number or other such symbol. What is missing is a known map between the physical output and the possible solutions to the problem.%, even if the map is merely the obvious map from the symbol ``$5$'' to the value 5.
%\footnote{There are voices that disagree and e.g.\ claim that the atoms in the wall of this room are constantly performing computationsREFERENCE, but that we simply do not have the power to utilize those. While everyone is free to define computation the way they want, I do not think that such a way of understanding computation is very fruitful as it greatly differs from the natural usage of the term.}

	The important subtlety here, it seems to me, is that the map $\mathfrak{I}$ must be \emph{known}, that is, it must be \emph{known to some agent running the computation}. In some sense (depending on what we mean by existence in the context of mathematical entities) all maps exist. However, for a process to be a computation (and thereby produce some kind of information at the end) the agent must know how to \emph{interpret} the physical outputs obtained, i.e.\ he or she must know $\mathfrak{I}$. Let us briefly digress and ask: How may one obtain knowledge of $\mathfrak{I}$? In order to answer this, consider the complete procedure necessary for computing the solution to a problem.
	
	Before the procedure may begin, there must be a well-defined problem for which a solution is sought. The problem may be an element of a whole class of problems of the same type, in which case it may be convenient to express it as the task to find the value of a function for a particular set of arguments. For example, evaluating ``$5+3$'' is a case of evaluating the function $f(x,y)=x+y$ for values $x=5$ and $y=3$. The problem is mapped onto the initial state of the system that is to execute the computation. The system then evolves according to some dynamics determined by the laws of physics.\footnote{A slightly different description would be to say that the evolution of the system is determined by the laws of physics and the state of apparatus, i.e.\ the set-up of gates (classical or quantum) in a circuit and so on. However, in our description the gates etc.\ are included in what we refer to as ``the system'' and their setup is therefore part of the initial state. Our description is more general in what kind of system may be performing the computation.}
At some point during this evolution the agent reads off some kind of output value from part of the system. For example, the state of the system may at some point be such that the symbol ``$8$'' is visible on a screen. At this point the agent applies his knowledge of the map $\mathfrak{I}$ to interpret the symbol ``$8$'' as the value 8, which forms the solution to the posed problem.

	The agent can infer the correct map $\mathfrak{I}$ from the map $\mathfrak{S}$ that translated the problem into the initial state of the system and some knowledge of physical laws that govern the dynamics.\footnote{Note though that in practice this rarely means that knowledge of the fundamental dynamics of the system is required. We do not need to know about atomic physics or even electronics to use a calculator. This is because others have done the work for us. The people who made the calculator constructed it in such a way that the possible outputs correspond to our familiar symbols for numbers and the map $\mathfrak{I}$ is (fortunately) the standard one. In fact, we are so used to this map that calculator manuals do not even tell us how to interpret the symbol ``5'' on the screen, although if the calculator has more advanced functions the manual often does specify how to set the initial conditions (that is, press the right combination of keys) in order to compute the right problem. Note that if the calculator has a fault and e.g.\ shows the symbol ``$5$'' when typing the ``$4$''-key and vice versa, it is still a perfectly good calculator -- we just have to change the map $\mathfrak{S}$, i.e. type ``$4$'',``$+$'',``$3$'' instead of ``$5$'',``$+$'',``$3$'' in order to compute the problem ``$5+3$'' and so on.}
	%Thus \emph{a priori} there are two ways to compute the solution to a given problem: First, one may take some system found in nature, look at the details of the dynamics that govern it and the initial state and infer the correct interpretation $\mathfrak{I}$ from some choice of $\mathfrak{S}$. In general, this is extremely hard and practically impossible. Most systems have dynamics that are far too complicated and we'd need a more powerful computer to compute the workings of these systems in the first place. In particular, it may require us to know the solution (or at least solutions to comparable problems) in advance and hence pre-empt the purpose of the computation.
	Thus to compute the solution to a given problem, we have to
	%The easier (and actually useful) way is to 
	construct a system in such a way that the maps $\mathfrak{S}$ and $\mathfrak{I}$ are simple and practically we also would like the system to allow a large number of different inputs.\footnote{e.g.\ a basic calculator capable of the operations addition, subtraction, multiplication and division would require three inputs $x,y,t$ where $t\in\{+,-,\cdot,\div\}$ determines the type of operation.} Such systems are called computers.
	
	We have intentionally avoided the term \emph{algorithm} so far. An algorithm is an abstract set of instructions to solve a computational problem. To ``run'' the algorithm these instructions have to be translated (via some suitable map) into steps in the dynamical evolution of the computer. Hence an algorithm is a useful tool to set up an appropriate system, yet it is not part of the physical description of the computation, although speaking in terms of algorithms can be extremely useful and we will do so too. The reason we often think of algorithms as something a computer actually does is that we are used to machines that can be prepared for computation through a very high level description, such as programming languages. Others have done most of the work for us and constructed devices that can be prepared in various useful initial states (such that the output map is easily interpretable) by merely pressing the right combination of keys. When we say that a machine executes an algorithm, we mean that the system that is the machine performs a computation corresponding to the algorithm in the described manner.

	The description so far has been general to include any type of (suitable) system and any kind of evolution. It applies equally to classical and quantum and perhaps other not yet discovered types of computers. Crucially, no matter what kind of laws govern the dynamics, an interpretable output is always required. This also means that the maximal possible amount of data a computer can compute is limited by the size of its output. A computer the size of the universe can still only compute a single bit of data if the interpretable output is limited to one bit. If one analyses the dynamics of the computer it might \emph{seem} as if various other bits of data are computed in the process before arriving at the final result. Note however that these are not physically accessible. They just appear in a mathematical description of the evolution. From this description it is in some cases easy to see how one could construct a very similar computer that yields those intermediate results, which explains the reasoning behind the claim that the given computer computes these. However, great care is necessary to distinguish cases in which this is not possible. I will return to this in the discussion of Deutsch's claim concerning computation in parallel universes.\\

		\subsection{The exponential speed-up of quantum computation}
		
	Consider now the difference between quantum and classical computers. A \emph{quantum computer} is a suitable physical system (a computer) whose dynamics include uniquely quantum mechanical effects. It is not merely that quantum theory must be invoked to provide an accurate description of the dynamics of the system (this is the case for any system whose components are small enough that quantum effects become relevant), but that the dynamics include evolutions that cannot be modelled or described classically. In particular, the description of the dynamics includes operations that cannot be modelled on a classical Turing machine.\footnote{Deutsch 1997, \cite{Deutsch1997}} 
	
	The employment of such quantum phenomena allows for an exponential speed-up in the computation of particular types of computational problems. In order to be more specific, first consider the notion of the size in the context of a computation. Consider a set of similar computational problems that only vary in the quantity of information they require as input. For example, factorising the numbers 20 and 2000 are two similar computational problems, but 20 requires only 5 bits of input (10100 in binary), while 2000 requires 11 bits (11111010000 in binary). Or consider searching a list of $n$ names for a particular entry. The problem requires different amounts of input data for different $n$ but the nature of the problem remains the same. To say that quantum computers provide an exponential speed-up compared to classical computers is to say that the resources required (time and memory) scale at most polynomially with the size of the problem, while in classical computing they scale exponentially. Quantifying the size of the problem by a parameter $n$, this implies that while there is no guarantee that for any particular $n$ the computation will be faster on a quantum than on a classical computer, there is a critical size $n_0$ such that for all $n>n_0$ the quantum computer is faster (and indeed the greater $n$ the greater the relative speed-up).
	
	If a computer (quantum or not) computes a certain problem with resources scaling polynomially with the size of the system, then we call the computation \emph{efficient}. Otherwise (i.e.\ if the resources scale exponentially), we say that the computation is \emph{inefficient}. Perhaps the main reason quantum computing has created interest is the possibility of factorising numbers efficiently and its implication for cryptography. An appropriate quantum algorithm (i.e.\ an abstract high level mathematical description of the required dynamics) was given by Shor in 1994.\footnote{1994, \cite{Shor1994}} %We will discuss exactly what aspects of quantum mechanics allow for the speed-up in section \ref{Conclusions} in the light of our analysis of quantum computing from the PWT point of view in section \ref{QCinPWT}.
	
	There are simple examples that illustrate the possible speed-up obtained through the employment of quantum effects. Unlike Shor's algorithm these examples may only be of limited practical use, but they do serve as good toy models. In particular, consider what has become known as Deutsch's algorithm. The problem was presented and a single bit solution given by Deutsch himself.\footnote{1985b, \cite{Deutsch1985b}} A more generalized n-bit solution was given by Deutsch and Josza in 1992\footnote{1992, \cite{DeutschJosza1992}} and has become known as the Deutsch-Josza algorithm. An improved solution has been given by Cleve, Ekert, Macchiavello and Mosca \footnote{1998, \cite{CEMM1998}}.\\

	\subsection{The Deutsch Algorithm}\label{DeutschAlgorithm}
	
	Here we will focus on the one-bit Deutsch algorithm as it already illustrates the use of quantum phenomena in order to perform computation with fewer resources than any classical computation would require. The posed problem is this: An ``oracle'' (that is, a black box that may form part of our circuit) contains dynamics that can be abstractly described as acting as one of the four possible functions
	$f_i: \{0,1\}\rightarrow\{0,1\}$
	\begin{eqnarray} 
		\notag &&f_0(0)=0;\qquad f_0(1)=0\\
		\notag &&f_1(0)=0;\qquad f_1(1)=1\\
		\notag &&f_2(0)=1;\qquad f_2(1)=0\\
		&&f_3(0)=1;\qquad f_3(1)=1
	\end{eqnarray}
	where 0 and 1 represent the physical states of some two state system.\footnote{Note that we have not yet specified what kind of states we are talking about. In quantum computing as described by orthodox quantum mechanics these states will be thought of as states in a Hilbert space, but oracles have been used in computer science even before the dawn of quantum computing.} We do not initially know which of the function $f_i$, $i\in\{0,1,2,3\}$, describes the action of the oracle, nor do we have any means of opening the box and study its dynamics. Our only hope of determining it is to incorporate the oracle into a circuit.
	
	Before continuing with the analysis of the Deutsch problem, let us review the terminology used here. Previously we have defined a \emph{computer} as a physical system that can be manipulated easily to take a large number of initial conditions (inputs) that lead to interpretable outputs and thereby provide solutions to computational problems if set up correctly. A \emph{circuit} is also such a system except that the dynamical evolution during the computation has been fixed and the only flexible part of the input is, speaking abstractly, an integer number of (classical or quantum) bits.\footnote{Sometimes the term ``circuit'' is used in the sense that even this input is already fixed.} Only these bits evolve during the computation. Thus a computer is ``programmable'' (can perform a variety of algorithms), a circuit is not.
	
	Return now to the Deutsch problem. Classically the oracle has to be called exactly twice to determine to which function $f_i$ it corresponds, namely once with a classical bit in the 0-state (i.e. once with a two-state system in the state we arbitrarily label ``0'') and once with a classical bit in the 1-state acting as argument for the function. The output is the value of $f_i$ and may be obtained by coupling the bit leaving the oracle to some suitable output device such as a screen or LED or dial.
	
	No quantum algorithm of orthodox quantum mechanics or an empirically equivalent theory such as equilibrium PWT can do any better. We do not have to know much about quantum computing to see why this is so. To specify which function represents the oracle, i.e.\ to specify the value of $i$ in $f_i$, \emph{two} bits are required, but the oracle only manipulates \emph{one} bit and what the rest of the circuit does to the bit before and after the oracle is invoked cannot possibly increase the amount of information. Beware though that this result is not trivial, since qubits (quantum bits) can be in a continuous number of superposition states (in the orthodox quantum mechanical description) and so can in theory store an infinite amount of information. The crucial point is that when the qubit is measured (in whatever way), that is, when an interpretable output is extracted, quantum mechanics limits this outcome to one of two states and in the process destroys the superposition (in the SQM description). Thus only a single bit of information may be obtained. We have shown in section \ref{PWT} that equilibrium PWT exactly agrees with standard quantum mechanics empirically and so the same result applies, although the description differs. Work by Valentini\footnote{2002, \cite{ValentiniSubquantumMeasurement}} relating to what he calls ``subquantum measurement'' has shown that the same cannot be said about non-equilibrium PWT because for certain non-equilibrium distributions $\rho\neq|\psi|^2$ there are possible measurements (i.e.\ couplings to an apparatus with an output device) that do not lead to an effective collapse and thus more than a single bit of information may be extracted.
		
		We have seen that quantum computing does not speed up just any computation. Two function calls are required both in classical and quantum computation due to the fundamental limitation on the amount of data obtained as output. 
		
		However, consider now a variation of the problem. Say instead of asking which function $f_i$ describes the oracle, we group the possible functions into pairs and ask which pair contains the function describing the oracle. For example, we may ask whether the correct function is in $\{f_0,f_1\}$ or in $\{f_2,f_3\}$. In this case, the answer is easily obtained with a single run of the circuit by setting the initial input (i.e. argument) bit to 0, since $f_i(0)=0\leftrightarrow f_i\in\{f_0,f_1\}$ and $f_i(0)=1\leftrightarrow f_i\in\{f_2,f_3\}$. The output is a single bit of information, sufficient to specify which of the pair contains the correct function. This applies to classical and quantum bits alike. A similar analysis is possible for the pairs $\{f_0,f_2\}$ and $\{f_1,f_3\}$ by setting the input bit to 1.
		
		The case is different for the pairing $\{f_0,f_3\}$ and $\{f_1,f_2\}$, i.e.\ for the question whether the function is constant ($f_0$ or $f_3$) or balanced ($f_1$ or $f_2$). The output is only one bit, yet there is no possible classical circuit that provides the required information with a single function call. However, there are quantum circuits that do the job. We will now describe the simplest (for all we know) of them. This is not quite the algorithm given originally by Deutsch.\footnote{1985b, \cite{Deutsch1985b}}
		
		We will now provide an abstract description of the Deutsch algorithm in terms of two-component vectors describing the state of the system, denoting n-qubit gates (i.e.\ primitive elements of the circuit) by $2^n\times2^n$ square matrices. If the world is indeed fundamentally governed by equilibrium PWT dynamics, then we have to understand this description as one in terms of an \emph{effective} theory, namely orthodox quantum mechanics. In anticipation of the PWT analysis of the Deutsch algorithm in section \ref{QCinPWT} we will also analyse the circuit more dynamically by finding the Hamiltonians corresponding to its primitive elements.
		
		For the Deutsch algorithm, we require three components in our circuit: The Hadamard gate, a single qubit measurement device and the oracle itself. We use the output of the measurement device to define a basis
\begin{eqnarray}\ket{0}=\twovector{1}{0},\qquad \ket{1}=\twovector{0}{1}\end{eqnarray}
and assume that the oracle acts as one of $f_i$ relative to those basis vectors. Equivalently, we could define the basis in terms of the action of the oracle and employ a measurement device that measures in that same basis. Thus the output device will give outputs `0' or `1' corresponding to states $\ket{0}$ and $\ket{1}$ respectively and if the oracle corresponds to $f_2$, say, then its dynamics will take $\ket{0}\rightarrow\ket{1}$ and $\ket{1}\rightarrow\ket{0}$. Its action on superpositions follows from the linearity of quantum mechanics. 

	The Hadamard gate $H$ is a single qubit unitary gate. Its action is given by
\begin{eqnarray} H = \overroottwo{1}\twomatrix{1}{1}{1}{-1} \end{eqnarray}
and is denoted by\\
%%$\qquad$\xymatrix@R=5pt@C=10pt{& \Had & \n}\\ 
\begin{centering}$\qquad\qquad\qquad$\includegraphics[width=60pt]{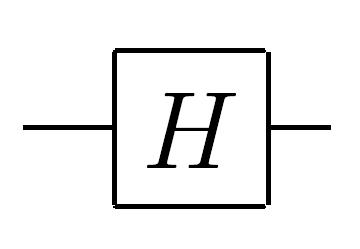}\end{centering}\\
We note that, as required by standard quantum mechanics, the evolution is unitary, as can be easily verified:
\begin{eqnarray} H\cdot H^{\dagger}=H\cdot H = \mathds{1}\end{eqnarray}

	We describe the oracle as a two-qubit gate $U_{f(x)}$ where $f(x)=f_i(x)$ for some $i\in\{0,1,2,3\}$, i.e.\ one of the four possible binary functions as described above. Writing $\ket{x}\ket{y}=\ket{x,y}$, the action of $U_{f(x)}$ is
	\begin{eqnarray}U_{f(x)}\ket{x,y}=\ket{x,y\oplus f(x)}\end{eqnarray}
where $\oplus$ stands for addition modulo 2. The effect of the oracle is then to leave the `data' qubit\footnote{terminology due to Nielsen \& Chuang, 2000,\cite{NielsenChuang2000}} unchanged and give the output of evaluating $f(x)$ via an auxiliary (`target') qubit. In particular, if the auxiliary qubit is initially in a state $\ket{y=0}$, then the combined output state would be $\ket{x,f(x)}$. Diagrammatically we represent this by\\
\begin{centering}$\qquad\qquad\qquad$\includegraphics[width=180pt]{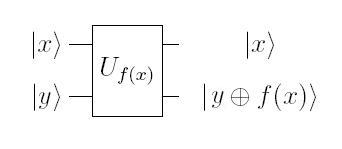}\end{centering}\\
%\def\Udata{\gnqubit{U_{f(x)}}{d}\w\A{Udata}}
%\def\Utarget{\gspace{U_{f(x)}}{d}\w\A{Utarget}}
%$\qquad$\xymatrix@R=5pt@C=10pt{\ket{x} & \Udata & \n & \ket{x} \\ 
%															 \ket{y} & \Utarget & \n & \ket{x\oplus f(x)}}\\
or, making the action of the oracle explicit in the form of a controlled NOT-gate, by\\
\begin{centering}$\qquad\qquad\qquad$\includegraphics[width=180pt]{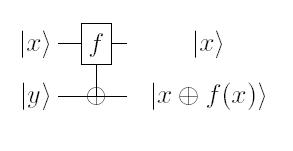}\end{centering}
%$\qquad$\xymatrix@R=5pt@C=10pt{\ket{x} & \op{f}\w\A{foracle} & \n & \ket{x} \\ 
%														   \ket{y} & \o\w\A{oracle} & \n & \ket{x\oplus f(x)}\\
%										\ar@{-}"oracle";"foracle"}.
															 
	Algebraically, we can write its evolution as
\begin{eqnarray} 
U_{f(x)} = {\ensuremath{\left[
	\begin{array}{cccc} \delta_{0,f(0)}& \delta_{1,f(0)}& 0& 0\\ 
											\delta_{1,f(0)}& \delta_{0,f(0)}& 0& 0\\
											0& 0& \delta_{0,f(1)}& \delta_{1,f(1)}\\
											0& 0& \delta_{1,f(1)}& \delta_{0,f(1)}\end{array}\right]}}
\end{eqnarray}
		where $\delta_{x,y}$ is the Kronecker-$\delta$ function and we interpret this to act on two-qubit state vectors written as 
	\begin{eqnarray}	\ket{0}\ket{0}=\fourvector{1}{0}{0}{0}, \qquad \ket{0}\ket{1}=\fourvector{0}{1}{0}{0}, \qquad \text{etc.} \end{eqnarray}
	 
	$U_{f(x)}$ is unitary as can easily be proved:
\begin{eqnarray}\notag U^{\dagger}\cdot U &=& U^2 \\
\notag &=& {\ensuremath{\left[
	\begin{array}{cccc} \delta_{0,f(0)}^2+\delta_{1,f(0)}^2 & 2\delta_{0,f(0)}\delta_{1,f(0)} & 0& 0\\ 
											2\delta_{0,f(0)}\delta_{1,f(0)} & \delta_{1,f(0)}^2+\delta_{0,f(0)}^2& 0& 0\\
											0& 0& \delta_{0,f(1)}^2+\delta_{1,f(1)}^2& 2\delta_{0,f(1)}\delta_{1,f(1)}\\
											0& 0& 2\delta_{0,f(1)}\delta_{1,f(1)} & \delta_{0,f(1)}^2+\delta_{1,f(1)}^2 \end{array}\right]}}\\
&=& \mathds{1}	\end{eqnarray}
since 
\begin{eqnarray}\notag \delta_{0,f(0)}\delta_{1,f(0)} &=& 0 \\ \delta_{0,f(0)}^2+\delta_{1,f(0)}^2 &=& 1 \end{eqnarray}
as $f(0)=0$ or $f(0)=1$ (but obviously never both) and similarly for $\delta$-functions involving $f(1)$. Thus unitarity has been proved.

	The final component of the circuit is the output device. In orthodox quantum mechanics this simply corresponds to a projective measurement (understood as a primitive notion as it appears in the postulates of the theory without definition in terms of previously defined entities). Since the measurement basis is the computational basis $\{\ket{0},\ket{1}\}$, we may describe the measurement by a set of operators $\{M_m\}$ as
\begin{eqnarray} M_0 = \outpr{0}{0} = \twomatrix{1}{0}{0}{0}, \qquad M_1=\outpr{1}{1} = \twomatrix{0}{0}{0}{1}. \end{eqnarray} 
According to the postulates of quantum mechanics the measurement yields an outcome 0 with probability $\inpr{\psi}{0}$ and an outcome 1 with probability $\inpr{\psi}{1}$. In PWT the description of measurement is more complicated since, as discussed in section \ref{PWT}, ``measurement'' is not a primitive notion. Details of a possible measurement dynamics for this particular case are found in section \ref{QCinPWT}. In the circuit diagram the measurement (together with any devices necessary to yield an interpretable output) is denoted by\\
\begin{centering}$\qquad\qquad\qquad$\includegraphics[width=60pt]{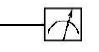}\end{centering}
%$\qquad$\xymatrix@R=5pt@C=10pt{& \n & \includegraphics[width=20pt]{Meter.jpg} }\\

We will now provide a description of the circuit in terms of the defined components. For an explanation of how to interpret quantum circuit diagrams, refer to e.g. \cite{NielsenChuang2000}. Moving from left to right through the diagram corresponds to moving forward in time. The final circuit diagram is\\
$\qquad\qquad\qquad$
%\xymatrix@R=5pt@C=10pt{\ket{0} & \n & \Had & \op{f}\w\A{foracle} & \Had & \n & \includegraphics[width=20pt]{Meter.jpg}\\ 
%											 \ket{1} & \n & \Had & \o\w\A{oracle} & \n & \n & \n .\\
%											 \ar@{-}"oracle";"foracle"}
\begin{centering}\includegraphics[width=240pt]{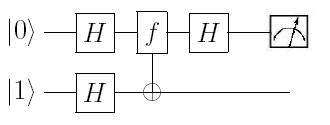}.\end{centering}
											 
Note that the input data and auxiliary qubits have been fixed to be $\ket{0}$ and $\ket{1}$ respectively. The one bit outcome of the measurement is sufficient to determine whether the function $f$ is constant or balanced.

To see this, define time steps such that $t=0$ initially, $t=1$ after the action of the first parallel pair of Hadamard gates, $t=2$ immediately after the action of the oracle and $t=3$ after the final Hadamard gate just before the measurement. Consider $\ket{\psi(t)}$ for $t=0,1,2,3$. Initially, the qubits are set up in the state
\begin{eqnarray}\ket{\psi(0)}=\ket{0}_d\ket{1}_a .\end{eqnarray}
The subscripts refer to the data and auxiliary qubit respectively. The Hadamard gates act individually on each qubit, maintaining a product state
\begin{eqnarray}\ket{\psi(1)}=[\overroottwo{1}(\ket{0}_d+\ket{1}_d)][\overroottwo{1}(\ket{0}_a-\ket{1}_a)]. \end{eqnarray}
The action of the oracle is to leave the data qubit unchanged in state $\ket{x}_d$ and act on the auxiliary qubit as $\ket{y}_a\rightarrow\ket{y\oplus f(x)}_a$. So the oracle acts as 
\begin{eqnarray}
\ket{x}_d[\overroottwo{1}(\ket{0}_a-\ket{1}_a)]\rightarrow(-1)^{f(x)}\ket{x}_d[\overroottwo{1}(\ket{0}_a-\ket{1}_a)],\end{eqnarray} i.e.\ the state remains unchanged up to a phase. Specifically we see that
\begin{eqnarray}\notag
&&\ket{0}_d[\overroottwo{1}(\ket{0}_a-\ket{1}_a)]\rightarrow(-1)^{f(0)}\ket{0}_d[\overroottwo{1}(\ket{0}_a-\ket{1}_a)] \\
&&\ket{1}_d[\overroottwo{1}(\ket{0}_a-\ket{1}_a)]\rightarrow(-1)^{f(1)}\ket{1}_d[\overroottwo{1}(\ket{0}_a-\ket{1}_a)]	\end{eqnarray}
and so by linearity
\begin{eqnarray}\notag &&\ket{\psi(1)}\rightarrow\ket{\psi(2)}=\pm[\overroottwo{1}(\ket{0}_d+\ket{1}_d)][\overroottwo{1}(\ket{0}_a-\ket{1}_a)] \qquad\mbox{ if } f(0)=f(1)\\ \\ \notag
&&\ket{\psi(1)}\rightarrow\ket{\psi(2)}=\pm[\overroottwo{1}(\ket{0}_d-\ket{1}_d)][\overroottwo{1}(\ket{0}_a-\ket{1}_a)] \qquad\mbox{ if } f(0)\neq f(1).\\	\end{eqnarray}
The final Hadamard gate acting on the data qubit then leads to the state
\begin{eqnarray}&&\ket{\psi(3)}=\pm\ket{0}_d[\overroottwo{1}(\ket{0}_a-\ket{1}_a)] \qquad\mbox{ if } f(0)=f(1)\\
&&\ket{\psi(3)}=\pm\ket{1}_d[\overroottwo{1}(\ket{0}_a-\ket{1}_a)] \qquad\mbox{ if } f(0)\neq f(1).	\end{eqnarray}
Measurement of the data qubit in the computational basis thus reveals if $f(0)=f(1)$ or $f(0)\neq f(1)$ and thus whether $f\in \{f_0,f_3\}$ or $f\in \{f_1,f_2\}$. This is how the Deutsch algorithm is generally presented in textbooks on the subject.

	Alternatively, instead of describing a step-by-step evolution of the state, we could have considered the total action of the circuit at once. The initial state is given by
\begin{eqnarray}\ket{\psi(0)}=\ket{0}_d\ket{1}_a=\fourvector{0}{1}{0}{0}. \end{eqnarray}
The total evolution operator of the Deutsch algorithm (without the measurement) is given by
\begin{eqnarray}\notag D &=& H\otimes \mathds{1} \cdot U_{f(x)} \cdot H\otimes H \\
									\notag &=& \overroottwo{1}{\ensuremath{\left[\begin{array}{cccc}
													1 & 0 & 1 & 0 \\
													0 & 1 & 0 & 1 \\
													1 & 0 & -1 & 0 \\
													0 & 1 & 0 & -1 \end{array}\right]}}\cdot													
												{\ensuremath{\left[\begin{array}{cccc} 
													\delta_{0,f(0)}& \delta_{1,f(0)}& 0& 0\\ 
													\delta_{1,f(0)}& \delta_{0,f(0)}& 0& 0\\
													0& 0& \delta_{0,f(1)}& \delta_{1,f(1)}\\
													0& 0& \delta_{1,f(1)}& \delta_{0,f(1)}\end{array}\right]}} \cdot
												\frac{1}{2}{\ensuremath{\left[\begin{array}{cccc}
													1 & 1 & 1 & 1 \\
													1 & -1 & 1 & -1 \\
													1 & 1 & -1 & -1 \\
													1 & -1 & -1 & 1 \end{array}\right]}}\\											
\end{eqnarray}
Multiplying the matrices explicitly, and using the fact that if $f$ is \emph{constant}, then
\begin{eqnarray} \delta_{0,f(0)}=\delta_{0,f(1)}, \qquad \delta_{1,f(0)}=\delta_{1,f(1)} \end{eqnarray}
and that if $f$ is \emph{balanced}, then
\begin{eqnarray} \delta_{0,f(0)}=1-\delta_{0,f(1)}, \qquad \delta_{1,f(0)}=1-\delta_{1,f(1)}, \end{eqnarray}
as well as that in both cases
\begin{eqnarray} \delta_{0,f(0)}+\delta_{1,f(0)}=1, \qquad \delta_{0,f(1)}+\delta_{1,f(1)}=1, \end{eqnarray}
we find
\begin{eqnarray}\notag
D_{constant} &=& \overroottwo{1} {\ensuremath{\left[\begin{array}{cccc}
													1 & \pm1 & 0 & 0 \\
													1 & \mp1 & 0 & 0 \\
													0 & 0 & 1 & \pm1 \\
													0 & 0 & 1 & \mp1 \end{array}\right]}} \\
D_{balanced} &=& \overroottwo{1}{\ensuremath{\left[\begin{array}{cccc}
													1 & 0 & 0 & \pm1 \\
													1 & 0 & 0 & \mp1\\
													0 & \pm1 & 1 & 0 \\
													0 & \mp1 & 1 & 0 \end{array}\right]}} \end{eqnarray}
where the subscripts refer to the evolution operators for constant and balanced oracle functions respectively and the ambiguity of sign come from the fact that there are two possible constant and two possible balanced functions.

Acting with $D_{constant}$ and $D_{balanced}$ on the initial state $\ket{\psi(0)}$ we find
\begin{eqnarray}\notag
\ket{\psi_{final}}_{constant} &&= D_{constant}\ket{\psi(0)} = \overroottwo{1} {\ensuremath{\left[\begin{array}{cccc}
													1 & \pm1 & 0 & 0 \\
													1 & \mp1 & 0 & 0 \\
													0 & 0 & 1 & \pm1 \\
													0 & 0 & 1 & \mp1 \end{array}\right]}} 
													\fourvector{0}{1}{0}{0} = \pm\overroottwo{1}\fourvector{1}{-1}{0}{0} \\ \\ \notag
\ket{\psi_{final}}_{balanced} &&= D_{balanced}\ket{\psi(0)} = \overroottwo{1}{\ensuremath{\left[\begin{array}{cccc}
													1 & 0 & 0 & \pm1 \\
													1 & 0 & 0 & \mp1\\
													0 & \pm1 & 1 & 0 \\
													0 & \mp1 & 1 & 0 \end{array}\right]}} 
													\fourvector{0}{1}{0}{0} = \pm\overroottwo{1}\fourvector{0}{0}{1}{-1}.\\
\end{eqnarray}
Here we already see that a measurement of the final state of the data qubit unambiguously reveals whether the evolution corresponded to a constant or balanced function $f$. Explicitly, the state of the subsystem that is the data qubit can be obtained by taking the partial trace (over the auxiliary qubit) of the density matrices $\rho_{constant/balanced}$ corresponding to the final states $\ket{\psi_{final}}_{constant/balanced}$:
\begin{eqnarray}\notag
\rho^{data}_{constant} &=& Tr_A(\rho_{constant}) = \frac{1}{2}Tr_A{\ensuremath{\left[\begin{array}{cccc}
													1 & -1 & 0 & 0 \\
													-1 & 1 & 0 & 0\\
													0 & 0 & 0 & 0 \\
													0 & 0 & 0 & 0 \end{array}\right]}} = \twomatrix{1}{0}{0}{0} = \ket{0}_d\bra{0}_d \\ \\ \notag
\rho^{data}_{balanced} &=& Tr_A(\rho_{balanced}) = \frac{1}{2}Tr_A{\ensuremath{\left[\begin{array}{cccc}
													0 & 0 & 0 & 0 \\
													0 & 0 & 0 & 0\\
													0 & 0 & 1 & -1 \\
													0 & 0 & -1 & 1 \end{array}\right]}} = \twomatrix{0}{0}{0}{1} = \ket{1}_d\bra{1}_d.\\
\end{eqnarray}
Thus the measurement in the computational basis provides the answer sought. This analysis is less suggestive of the fallacy that both $f(0)$ and $f(1)$ are ever calculated. 

 As a heuristic aside, note that \emph{prima facie} it seems counterintuitive to think that a measurement of the data qubit should reveal anything. We have noted that $H^2=\mathds{1}$ and that the oracle does not change the data qubit and so we might naively expect that no matter what we should get outcome 0. However, this is obviously not true. The mistake here is to forget that while the oracle does indeed not change the state of the data qubit directly, it does however in general entangle the two qubits, e.g.\
 	\begin{eqnarray} \frac{1}{\sqrt{2}}(\ket{0}_d+\ket{1}_d)\ket{0}_a \quad \underset{f_1}{\rightarrow} \quad \frac{1}{\sqrt{2}}(\ket{0}_d\ket{0}_a + \ket{1}_d\ket{1}_a). \end{eqnarray}
Hence after the operation of the oracle we should only think of the qubits as a single composite system. Interestingly, if the second qubit is in the state $\frac{1}{\sqrt{2}}(\ket{0}_a-\ket{1}_a)$ (as it is for the Deutsch algorithm), the resultant state is actually separable! This is an ``accidental'' peculiarity of this particular state only. It does not warrant a simpler analysis than we would have had to do for other states. This is also a good example illustrating how entanglement is still not entirely understood. Through a mechanical mathematical analysis we discover that the final state is a product state only if the initial state of the auxiliary qubit is the state above, but to understand this conceptually remains difficult. 
 
 Before continuing with a more dynamical analysis of the circuit, we note that the Deutsch algorithm is a good example illustrating the fact that quantum computing only allows for a speed-up in a small number of apparently contrived scenarios. Here only the question whether the function $f$ describing the action of the oracle is in $\{f_0,f_3\}$ or $\{f_1,f_2\}$ may be answered with fewer resources. In the other cases (i.e. trying to determine $f=f_i$ for which $i\in \{0,1,2,3\}$, or any other pairing) no quantum algorithm is known that is faster than the best classical ones. The challenge of quantum computer scientists is then to find ways of utilising computations that can be accelerated using quantum circuits. A prime example of this is Shor's algorithm and the insight that a fast way of period finding can be used for methods of efficiently factorising large integers.\footnote{See Mermin 2007b, \cite{Mermin2007b}}

	Quantum circuit diagrams are diagrammatical representations of how the state of an evolving subsystem (the qubits) interacts with the rest of the computer (the gates, measurement devices, etc.). So far our description has been abstractly formulated in the language of a general two-state quantum system whose states can be written as two-component vectors. That is, the description has been purely in terms of states of a $2^n$-dimensional Hilbert space, where $n$ is the number of qubits whose evolution is relevant to the computation. Hence the description has been entirely in terms of mathematical entities. It is also the type of description suitable for a treatment of quantum computing based on orthodox quantum theory.
	
	For PWT the discussion so far is useful in as far as orthodox quantum mechanics is an effective theory of equilibrium PWT, as well as a starting point for a more dynamical description in terms of the type of interaction Hamiltonians required. For a fully \emph{physical} description we first need to know exactly what kind of system constitutes the computer, i.e.\ how the qubits and their manipulation is implemented. In order to calculate the trajectory of the particle in PWT such a treatment is necessary and is attempted in section 4 for two (if somewhat unrealistically simplified) specific implementations of qubits. The dynamical description given here will still be in terms of states represented by two-vectors.

	The primitive elements of the circuit represent certain evolutions of the qubit	due to the chosen setup of the computer. For the implementation of the Deutsch algorithm presented here, at least in the context of orthodox quantum mechanics, only two constituents require description:\ the Hadamard gate and the oracle. Measurement in the context of PWT has been described in section \ref{PWT} and we will see its use in section \ref{QCinPWT}.
	
	A word on notation: Unfortunately both the Hadamard gate and the Hamiltonian are almost universally referred to as `$H$'; we do not wish to deviate from this standard notation. In order to avoid ambiguity we will at all times equip the symbol for the Hamiltonian with a subscript indicating the system whose dynamics the Hamiltonian encodes. An `$H$' without a subscript will always refer to the evolution matrix of the Hadamard gate.
	
	The operators representing the evolution of a quantum mechanical system are unitary. In discrete systems such as a set of qubits the evolution may be represented as a unitary matrix $U$, as we have done previously. The evolution is generated by a Hermitian\footnote{The last ten or so years have brought the discovery of non-Hermitian PT-symmetric Hamiltonians in quantum mechanics and the field is now at a stage where first experimental results may be obtained. While a fascinating new area of research, we do not have the space to treat it here.} matrix $G$, that is
	\begin{eqnarray} U = e^{iaG} \end{eqnarray}
where the factor $i$ in the exponent is a choice of convention and $a$ an appropriate parameter. In particular, for a system satisfying the Schr\"odinger equation
\begin{equation}
i\frac{\partial}{\partial t}\ket{\psi} = H_{sys}\ket{\psi},
\end{equation}
where $H_{sys}$ is the Hamiltonian of the system, we obtain the general solution
\begin{eqnarray} U = e^{-iH_{sys}t} \end{eqnarray}
as can be verified by substituting $\ket{\psi(t)}=e^{-iH_{sys}(t-t_0)}\ket{\psi(t_0)}$ into the Schr\"odinger equation. The task now is to find $H_{Had}$ and $H_{oracle}$ that generate the evolution operators $H$ and $U_{f(x)}$ respectively.\footnote{There is a certain inconsistency here in the use of the term ``generate''. Mathematically we said that $G$ generated $U=e^{iaG}$. Now we say that the Hamiltonian $H_{sys}$ generates the evolution $U=e^{-iH_{sys}t}$. However, this slightly ambiguous use seems to be common in the literature (``the Hamiltonian generates the time evolution of the system'') and we will not deviate from it here as it is clear from the context what is meant.} 

	Let $t_0$ denote the moment the interaction Hamiltonian is ``switched on'' and $t_1$ the moment it is ``switched off''. In other words, let $T=t_1-t_0$ be the time of the interaction, i.e.\ the time for the qubit to pass through the particular considered gate. We assume here that $T$ is short compared to the natural evolution of the qubit, so we can ignore terms other than the chosen interaction (represented by the evolution matrix) of the gate and qubit.
	
	Note that there is a certain freedom of parameter choice: From the evolution matrices we will be able to calculate the required exponents $-iH_{Had}T$ and $-iH_{oracle}T$ to achieve this evolution. This means that we can scale the Hamiltonian by some real factor as long as we adjust the interaction time $T$ accordingly (provided $T$ is still short enough that we can ignore the free evolution of the qubits). In reality the choice will be limited by physical constraints on the type of  system used as computer.
	
	Let us denote the Pauli matrices by
\begin{eqnarray} X=\twomatrix{0}{1}{1}{0}, \qquad Y=\twomatrix{0}{-i}{i}{0}, \qquad Z=\twomatrix{1}{0}{0}{-1} \end{eqnarray}
and let $\vec{\sigma}$ denote the ``vector'' $(X,Y,Z)$. Let us define the rotation operators
\begin{eqnarray}\label{rotations}
R_x(\theta)\notag &&= e^{-i\theta X/2} = \cos(\frac{\theta}{2})\mathds{1}-i\sin(\frac{\theta}{2})X = \twomatrix{\cos(\frac{\theta}{2})}{-i\sin(\frac{\theta}{2})}{-i\sin(\frac{\theta}{2})}{cos(\frac{\theta}{2})} \\
R_y(\theta)\notag &&= e^{-i\theta Y/2} = \cos(\frac{\theta}{2})\mathds{1}-i\sin(\frac{\theta}{2})Y = \twomatrix{\cos(\frac{\theta}{2})}{-\sin(\frac{\theta}{2})}{\sin(\frac{\theta}{2})}{\cos(\frac{\theta}{2})} \\
R_z(\theta) &&= e^{-i\theta Z/2} = \cos(\frac{\theta}{2})\mathds{1}-i\sin(\frac{\theta}{2})Z = \twomatrix{e^{-i\theta/2}}{0}{0}{e^{i\theta/2}},
\end{eqnarray}
where the matrix form may be obtained from expanding the integral and using $X^2=Y^2=Z^2=\mathds{1}$. They are called thus as their action on a state in the Hilbert space corresponds to the rotation of the Bloch vector of that state around the $x$, $y$ and $z$ axes by an angle $\theta$. We generalise this to rotations around any axis given by a unit vector $\vec{n}$:
\begin{eqnarray}R_{\vec{n}}(\theta) = e^{-i\theta \vec{n}\cdot\vec{\sigma}/2} = \cos(\frac{\theta}{2})\mathds{1}-i\sin(\frac{\theta}{2})(n_xX+n_yY+n_zZ) \end{eqnarray}
Any single qubit unitary operator may then be written in the form
\begin{eqnarray} U = e^{i\alpha}R_{\vec{n}}(\theta). \end{eqnarray}
We may see this geometrically using the correspondence of the Hilbert space with the Bloch sphere: A pure state up to an overall phase is given by a state on the surface of the Bloch sphere and any other point on the surface can be reached with an appropriate rotation $R_{\vec{n}}$. $e^{i\alpha}$ then specifies the overall phase.

	We now wish to find $\alpha$, $\vec{n}$ and $\theta$ for the Hadamard gate. Writing the general expression for $U$ in matrix form and equating it to the Hadamard gate, we require
\begin{eqnarray}\notag H = \overroottwo{1}\twomatrix{1}{1}{1}{-1} \underset{req}{=} e^{i\alpha}\twomatrix{\cos(\frac{\theta}{2})-in_z\sin(\frac{\theta}{2})}{-in_x\sin(\frac{\theta}{2})-n_y\sin(\frac{\theta}{2})}{-in_x\sin(\frac{\theta}{2})+n_y\sin(\frac{\theta}{2})}{\cos(\frac{\theta}{2})+in_z\sin(\frac{\theta}{2})}.\\ \end{eqnarray}
We can now solve for $\alpha$, $\vec{n}$ and $\theta$ by inspection. Note that the top left and bottom right entries of $H$ differ only by a sign, from which we infer that $\cos(\frac{\theta}{2})=0$ and so $\theta=\pi$, leaving us with
\begin{eqnarray}\overroottwo{1}\twomatrix{1}{1}{1}{-1} \underset{req}{=} e^{i\alpha}\twomatrix{-in_z}{-in_x-in_y}{-in_x+in_y}{in_z}. \end{eqnarray}
Noting further that the top right and bottom left entries are equal, we require $n_y=0$. Since all entries have the same magnitude, we have $n_x=n_z$ and by normalization of $\vec{n}$ this implies $n_x=n_z=1/\sqrt{2}$, i.e.\ \begin{eqnarray}\overroottwo{1}\twomatrix{1}{1}{1}{-1} =_{req} \overroottwo{1} e^{i\alpha}\twomatrix{-i}{-i}{-i}{i}. \end{eqnarray}
The matrix on the right-hand side is off target by a factor $i$ and thus the required phase factor is $\alpha=\pi/2$.

	In summary, we have obtained an expression of the Hadamard gate in the form of exponentials only, namely
\begin{eqnarray} H = e^{i\frac{\pi}{2}}e^{-i\frac{\pi}{2}\overroottwo{1}(X+Z)} = e^{-i\frac{\pi}{2}(\overroottwo{1}(X+Z)-\mathds{1})} = e^{-i\frac{\pi}{2}(H-\mathds{1})} \end{eqnarray}
using the fact that any matrix commutes with the identity.
Comparing this to the expression for the evolution as generated by the Hamiltonian $H_{Had}$, we find that
\begin{eqnarray} e^{-iH_{Had}T} = e^{-i\frac{\pi}{2}(H-\mathds{1})} \end{eqnarray}
and hence
\begin{eqnarray} H_{Had}T = \frac{\pi}{2}[H-\mathds{1}] %= \frac{\pi}{2}\twomatrix{-1+\overroottwo{1}}{\overroottwo{1}}{\overroottwo{1}}{-1-\overroottwo{1}}
.\end{eqnarray}
We have found the Hamiltonian required for implementing a Hadamard gate in terms of a general two-state system.

	We will now investigate the evolution $U_{f(x)}$ of the oracle. We had the general expression
\begin{eqnarray} 
U_{f(x)} = {\ensuremath{\left[
	\begin{array}{cccc} \delta_{0,f(0)}& \delta_{1,f(0)}& 0& 0\\ 
											\delta_{1,f(0)}& \delta_{0,f(0)}& 0& 0\\
											0& 0& \delta_{0,f(1)}& \delta_{1,f(1)}\\
											0& 0& \delta_{1,f(1)}& \delta_{0,f(1)}\end{array}\right]}},
\end{eqnarray}	
which we may write compactly in 2x2 block form as
\begin{eqnarray} U_{f(x)} = \twomatrix{\delta_{0,f(0)}\mathds{1}+\delta_{1,f(0)}X}{0}{0}{\delta_{0,f(1)}\mathds{1}+\delta_{1,f(1)}X}.
\end{eqnarray}
The fact that this expression is block-diagonal implies that we may solve for the generator of each block separately. We observe that 
\begin{eqnarray}\mathds{1} = e^{\mathds{O}}\end{eqnarray}
where $\mathds{O}$ refers to the zero 2x2 matrix.

	To find an expression for $X$ we recall the relation
\begin{eqnarray}
R_x(\theta) = \twomatrix{\cos(\frac{\theta}{2})}{-i\sin(\frac{\theta}{2})}{-i\sin(\frac{\theta}{2})}{\cos(\frac{\theta}{2})}
\end{eqnarray}
and choose $\theta=\pi$, yielding
\begin{eqnarray} R_x(\pi) = \twomatrix{0}{-i}{-i}{0} = -iX. \end{eqnarray}
Hence
\begin{eqnarray} X = iR_x(\pi) = e^{i\frac{\pi}{2}}R_x(\pi) = e^{i\frac{\pi}{2}}e^{-i\frac{\pi}{2}X} = e^{-i\frac{\pi}{2}(X-\mathds{1})}. \end{eqnarray}
For
\begin{eqnarray} U_{f(x)} = e^{-iH_{oracle}T} \end{eqnarray}
we thus find
\begin{eqnarray} H_{oracle}T = \twomatrix{\delta_{1,f(0)}\cdot\frac{\pi}{2}(X-\mathds{1})}{0}{0}{\delta_{1,f(1)}\cdot\frac{\pi}{2}(X-\mathds{1})}. \end{eqnarray}
		
		The dynamics of the whole circuit is then described by acting with $H_{Had}$ and $H_{oracle}$ for the right durations $T$ in the right order. This completes our discussion of the Deutsch algorithm in terms of Hilbert space states for a general two-state system in the language of orthodox quantum mechanics. In section \ref{QCinPWT} we will return to the algorithm with a more physical approach and its description in PWT.
		
		The more general Deutsch-Josza algorithm is a multi-qubit extensions of the one presented here and the task is to find whether a function $f:\{0,1\}^n\rightarrow\{0,1\}$ is constant or exactly balanced (suppose we know that these are the only two options).\footnote{A good overview of the Deutsch-Josza algorithm may be found in \cite{NielsenChuang2000}} An interesting feature of the general Deutsch-Josza algorithm is that at the end of the procedure $n$ one-qubit measurements are performed, theoretically yielding $n$ bits of information. However the information obtained is only one bit (constant or balanced). One might wonder if a better algorithm could be found that can provide more information from a single oracle call, perhaps one that does not require the promise that the function $f$ is constant or exactly balanced. The fact that the existing algorithm has to rely on this promise once again suggests how only very specific tasks experience a speed-up in quantum computing.\\

	\subsection{Physical Implementations}
	
	This is not the place for a detailed review of physical implementations of quantum computer. In section \ref{QCinPWT} we will consider two somewhat unrealistic models that serve well in an analysis of how quantum algorithms may be understood in principle in terms of PWT. However, a few words regarding the state of the art are in order, if only to place the rest of the discussion into context.
	
	Experimentalists are nowhere near constructing any type of quantum computer that might have a purpose beyond illustrating the possibility of quantum computation through very simple algorithms. Any type of system that is to serve as a quantum computer must satisfy at least two requirements. Firstly, it must provide a stable way to encode qubits. That is, decoherence times must be significantly larger than computation times. Secondly, the qubits must be accessible. The experimentalist must be able to initially prepare fiducial states for the purposes of computation, manipulate the qubits\footnote{It may be sufficient to have control over only a few of the qubits and use a fixed coupling between them to control other qubits via the control of those few. See e.g.\ \cite{BurgarthEtAl2008} and \cite{BurgarthEtAl2009}} by being able to implement a complete set of qubit gates (that is, a set of qubit gates from which any other gate can be constructed --- this can, for example, be done with only the CNOT-gate and single qubit gates) and finally perform reliable measurements on at least some of the qubits. These two types of condition can be conflicting: On the one hand we require the qubits to be sufficiently isolated from their environments in order to be stable. On the other hand we require physical access for preparation, control and measurement. In other words, the system that constitutes the qubits must be well isolated except for very controlled interactions with the experimentalist's apparatus. To achieve this is the challenge of experimental quantum computing.
	
	It is unlikely that any single type of system will be found that fulfils these criteria. A hybrid computer, where qubits may be stored in one form, then converted into another for manipulation, for example, is more likely to be successful. Ion trap quantum computers of about seven to ten qubits have been constructed, but scaling them up to larger systems remains difficult. In solid state quantum computers maintaining control for more than two or so qubits continues to be challenging. Each of these comprises its own entire field of research and a detailed discussion is beyond the scope of this work. 
	
	The examples considered in the next section in terms of energy states of an infinite well and in terms of a simple spin description are merely toy models. However, they do illustrate in principle how the abstract description above translates into the manipulation of physical systems and, for the purposes of this work in particular, they serve as simple examples illustrating how PWT accounts for the effects achieved through quantum computation.

%______________________________________________________________________________________________________________________		
		
	\pagebreak		
	\section{Quantum Computing from the Pilot-Wave viewpoint} \label{QCinPWT}
			
	\begin{quote} ``[O]ne must warn young scientists against spending their time on calculating Bohm trajectories ... They would be investigating mere phantoms.'' (Zeh (1999), p. 199)
	\end{quote}
	In this section we will calculate the trajectories of model systems capable of implementing the Deutsch algorithm. However, we already know the result. In section \ref{PWT} we showed that PWT reproduces any observable result of SQM. A quantum computer is just a specific physical system and so this result applies here too. We thus obtain the same computational output in a world fundamentally ruled by equilibrium PWT as in one ruled by SQM or MWT. The calculations here are then merely a technical exercise. However, firstly such exercises help us develop a more intuitive understanding of the theory in question and secondly we thereby hope to gain some insight in the evolution that constitutes the computation in the PWT picture, which will be useful in answering the (very vague) question of what it is physically that explains the difference in computational speed between classical and quantum computing.
	
	In section \ref{Denial} I have already stated my disagreement with Zeh. However, even if PWT is no more than a mathematical model with little ontological resemblence to reality and we are indeed calculating nothing but ``phantoms'', two points still stand: (1) The fact that a coherent description of observable phenomena using real trajectories is at all possible, even if ultimately untrue, shows that Zeh and others in his camp have no proof that these trajectories are indeed nothing but phantoms, and (2) it thus follows that Deutsch's claim that electron diffraction and other quantum effects (and hence also quantum \emph{computing} effects) can only be explained coherently by a many-world theory (\emph{ergo} MWT be true) is simply false. Also, we might consider these phantoms pedagogically useful.\\
	
	\subsection{Bell's spin toy model}
	
Let us first analyse the Deutsch algorithm as it might be implemented by Bell's toy model for spin as described in section \ref{Spin}. The algorithm uses two qubits and hence we need two spin-$\frac{1}{2}$ particles, in addition to a single measurement device, characterised by a variable $y$, that is required for the final step of the algorithm. Recall that initially the wave function is a narrow wave packed $\phi_0(y)$ centred at $y=0$. The total initial wave function is then given by
	\begin{eqnarray} \psi_{mn}(y,t_0) = \phi_0(y)d_m(t_0)a_n(t_0), \end{eqnarray}
where we define $d_m(t)$ and $a_n(t)$ to be the wave functions of the data and auxiliary qubit respectively and $t_0$ is the point in time when the computation is set into motion. Calculating the evolution of the pilot-wave is then almost trivial as we have done nearly all of the work already. Given that the wave functions of the two particles are zero-dimensional and completely specified by a two-component vector of unit length, the two states of the abstract description provided in the previous section can be directly identified with the two eigenstates of $Z$:
	\begin{eqnarray} \notag &&\ket{0}_d \leftrightarrow d_m = \twovector{1}{0}_m, \qquad \ket{1}_d \leftrightarrow d_m = \twovector{0}{1}_m, \\
	&&\ket{0}_a \leftrightarrow a_n = \twovector{1}{0}_n, \qquad \ket{1}_a \leftrightarrow a_n = \twovector{0}{1}_n.
	\end{eqnarray}
The evolution induced by the two Hadamard gates is given by
	\begin{eqnarray} \phi(y)d_m(t_0)a_n(t_0) \rightarrow H_{mp}H_{nq} \phi_0(y)d_p(t_0)a_q(t_0) \end{eqnarray}
and for input states $d_m = \twovector{1}{0}_m$ and $a_n = \twovector{0}{1}_n$ this corresponds to 
	\begin{eqnarray} \phi_0(y)\twovector{1}{0}_m\twovector{0}{1}_n \rightarrow \frac{1}{2} \phi_0(y) \twovector{1}{1}_m\twovector{1}{-1}_n, \end{eqnarray}
which we choose to rewrite using pair labels in a single four component vector, i.e.\ as
	\begin{eqnarray} \frac{1}{2} \phi_0(y) \fourvector{1}{-1}{1}{-1}_{(mn)} \end{eqnarray}
with $m,n\in\{0,1\}$.
	
The action of the oracle is then
	\begin{eqnarray} \frac{1}{2} \phi_0(y) \fourvector{1}{-1}{1}{-1}_{(mn)} \rightarrow \frac{1}{2}\phi_0(y) U_{f(x)(mn)(pq)} \fourvector{1}{-1}{1}{-1}_{(pq)}. \end{eqnarray}
	
We have already worked through this evolution for the various possibilities for $f(x)$ in section \ref{DeutschAlgorithm} and the abstract results in terms of a four-dimensional complex vector space (essentially the Hilbert space of the SQM-description) translate straightforwardly into the evolution of the four-component pilot-wave in configuration space. Thus we find that the pilot-wave just prior to the final measurement is given by
	\begin{eqnarray}
	\notag &&\psi_{mn}(y,t_{pre-meas})=\phi_0(y) \frac{1}{\sqrt{2}}\fourvector{1}{-1}{0}{0}_{mn}	\qquad \text{if } f(0)=f(1)\\
			&&\psi_{mn}(y,t_{pre-meas})=\phi_0(y) \frac{1}{\sqrt{2}}\fourvector{0}{0}{1}{-1}_{mn}	\qquad \text{if } f(0)\neq f(1).
	\end{eqnarray}
We note that up to this point the qubits have not interacted with the measurement apparatus, whose wave function $\phi(y)$ is still unchanged. Hence the ensemble equilibrium distribution is unchanged too. 

	The interaction Hamiltonian for the subsequent measurement is given by 
	\begin{eqnarray} \hat{H}_{Bell} = -ig(Z\otimes \mathds{1})\frac{\partial}{\partial y} \end{eqnarray}
and so analogously to our analysis in section \ref{Spin} we obtain
	\begin{eqnarray} \notag &&\psi_{mn}(y,t_{post-meas}) = \psi_{mn}(y-g\Delta t,t_{pre-meas})	\qquad \text{if } f(0)=f(1)\\
									 &&\psi_{mn}(y,t_{post-meas}) = \psi_{mn}(y+g\Delta t,t_{pre-meas}) \qquad \text{if } f(0)\neq f(1)
					\end{eqnarray}
where $\Delta t = t_{post-meas}-t_{pre-meas}$. Thus, if the initial wave packet of the pointer is sufficiently narrow, we can unambiguously read off whether $f$ is constant or balanced. We note that this result does not strictly rely on being in quantum equilibrium, but any ensemble distribution that evolves in such a way that the two packets of non-zero density separate with the separation of the pilot-wave packets will do.
	
	If we wish we can also trace the evolution of the pilot-wave during the times ``inside'' a gate. For example, for the Hadamard gate we found the Hamiltonian
	\begin{eqnarray}H_{Had}T = \frac{\pi}{2}[H-\mathds{1}]. \end{eqnarray}
Setting $t=0$ when the gate is ``switched on'', we may consider the state of the wave at times $t=aT$ where $a\in[0,1]$. The evolution operator is 
	\begin{eqnarray}\notag U = e^{-iH_{Had}aT} &=& e^{-i\frac{a\pi}{2}(H-\mathds{1})} \\ \notag
									 &=& e^{-i\frac{a\pi}{2}H}e^{i\frac{a\pi}{2}\mathds{1}} \qquad\qquad\qquad\qquad \text{since } [H,\mathds{1}]=0 \\
									 &=& \left[\cos(\frac{a\pi}{2})\mathds{1}-i\sin(\frac{a\pi}{2})H\right]e^{i\frac{a\pi}{2}}.
		\end{eqnarray}
For $a=1$ this reduces to the complete Hadamard gate and for $a=0$ to the identity. The wave function of the pointer (and hence the ensemble distribution) is naturally unchanged while passing through the gate in this model for spin. As such, investigating the evolution ``within'' a gate is only of marginal interest here.

	We have discussed the evolution of the ensemble (unchanged until the measurement, then moving with the shift of the wave packet of the pointer) but have not yet calculated any actual trajectory. Prior to the final measurement the wave function of the pointer and that of the qubits is entirely independent. Hence for this period $(t<t_{measurement})$ we can write the total wave function as
	\begin{eqnarray} \psi(y,t) = \phi(y,t)r(t) \end{eqnarray}
where $r(t)$ is the normalised four-component zero-dimensional wave (i.e.\ $r(t)\in\mathds{C}^4$ with $|r| = 1$) of the two qubits and $\phi(y,t)$ is the wave function of the pointer, such that, for $t<t_{measurement}$, $\phi(y,t)=\phi_0(y)$. The Hamiltonian $H_{Gate}$ of any of the gates acts only on $r(t)$ and is expressible as a Hermitian matrix. Hence for the current we find unsurprisingly
	\begin{eqnarray} j(y,t)\notag &=& -\frac{\partial}{\partial t}|\psi(y,t)|^2 \\
												 \notag &=& -\left[\left(\frac{\partial}{\partial t}\psi(y,t)\right)^{\dagger}\psi(y,t)
												  		 				+\psi(y,t)\left(\frac{\partial}{\partial t}\psi(y,t)\right)\right] \\
												 \notag &=& -\left[(iH_{Gate}\psi(y,t))^{\dagger}\psi(y,t)+i\psi^{\dagger}(y,t)H_{Gate}\psi(y,t)\right] \\
												 \notag &=& i|\phi(y,t)|^2\left[r^{\dagger}(t)H^{\dagger}_{Gate}r(t)-r^{\dagger}(t)H_{Gate}r(t)\right] \\
												  &=& 0 \qquad \qquad \text{since } H_{Gate} \text{ is Hermitian}
	\end{eqnarray}
and so if we continue to neglect the free evolution of the pilot-wave, the equation of motion for $t<t_{measurement}$ is simply
	\begin{eqnarray} \frac{dy}{dt} = 0. \end{eqnarray}

	For the time of the measurement process we have already calculated the equation of motion (the existence of a second qubit is hereby irrelevant since the measurement interaction is only concerned with the first) and found (equation \ref{BellToyEOM})
	\begin{eqnarray}\frac{dy}{dt} = \pm g. \end{eqnarray}
So not only the wave packet and density distribution as a whole moves with velocity $g$ in the positive or negative $y$-direction depending on the oracle function $f$, but in fact the configuration (the particle/corpuscle) itself moves with this velocity. The trajectories of the ensemble are exactly parallel. The direction of the trajectories depend on the wave function. From this we may wish to conclude that most of the ``work'' of the computation is done by the wave function and the actual trajectory has little to do with it.\\

	\subsection{An infinite well model}\label{wellmodel}
	
	We will now consider a model in which qubits are not realised by the two spin states of spin-$\frac{1}{2}$ particles but the ground state and first excited state of particles in an infinite well, colloquially also known as particles in a box. In particular, to keep the description as simple as possible we will represent qubits by the states of a 1-dimensional well. Since the Deutsch algorithm requires two qubits, we will either need two such particle-well systems, or a single 2-dimensional one. While we will use the language of two distinct wells for the most part, implementing the coupling between the degrees of freedom describing the qubits as required for the oracle may actually be easier for the single 2-dimensional well.
	
	For further simplicity, assume that the box has a length $L = 1$ and is described by coordinates $x\in[0,1]$. The ground and first excited state in the position basis are then given by
	\begin{eqnarray} \inpr{x}{0} = \inpr{x}{\psi_1} = \sqrt{2}\sin(\pi x), \qquad\qquad \inpr{x}{1} = \inpr{x}{\psi_2} = \sqrt{2}\sin(2\pi x). \end{eqnarray}
Note that the discrepancy of labelling between the wave function $(\psi_1,\psi_2)$ and the state kets $(\ket{0},\ket{1})$ arises from conventions used in the general literature. Since these are energy eigenstates, their free time evolution is given by 
	\begin{eqnarray} \ket{\psi_n(t)} = e^{-iE_nt}\ket{\psi_n(0)} \qquad\qquad \text{with } E_n = \frac{1}{2}n^2\pi^2m \end{eqnarray}
and a general state $\ket{\psi} = a\ket{\psi_1} + b\ket{\psi_2}$ evolves into
	\begin{eqnarray} \ket{\psi(t)} &=& \notag e^{-iE_1t}a\ket{\psi_1} + e^{-iE_2t}b\ket{\psi_2} \\
																 &=& e^{-i(E_1+E_2)/2t}(ae^{-i\omega t}\ket{\psi_1}+be^{i\omega t}\ket{\psi_2}) \end{eqnarray}
with $\omega = \frac{1}{2}(E_1-E_2)$, which in abstract vector notation corresponds to a state $\twovector{a}{b}$ evolving according to a Hamiltonian $H_{free}=\omega Z$ and so its time evolution is $\twovector{a}{b}\rightarrow R_z(2\omega t)\twovector{a}{b}$, where $R_z(\theta) = e^{-i\theta Z/2}$ as defined in equation \ref{rotations}, plus an overall phase rotation. This means we can implement a $Z$-gate, for example, by simply waiting for the correct amount of time $\omega t=\pi$.

	We note that any single-qubit gate can be written in the form 
	\begin{eqnarray} U_{Gate} = e^{i\alpha}R_z(\beta)R_x(\gamma)R_z(\delta) \end{eqnarray}
	as may be easily seen by recalling that $R_x(\theta)$ and $R_z(\chi)$ correspond to rotations in the Bloch sphere by angles $\theta$ and $\chi$ around the x and z-axis respectively and that geometrically any arbitrary rotation may be implemented by three sequential rotations round these axes. $e^{i\alpha}$ fixes the overall phase, which in our example may be implemented using the total phase time evolution.
	
	Hence in order to be able to implement an arbitrary single-qubit gate the only task remaining is to find a way of performing a rotation $R_x(\theta)$ in our given model. This may be achieved by perturbing the potential (and hence the Hamiltonian) inside the well. In particular, adding the term\footnote{See e.g. Nielsen \& Chuang (2000), \cite{NielsenChuang2000}}
	\begin{eqnarray} \delta V(x) = -\frac{9\pi^2}{16}\left(x-\frac{1}{2}\right) \end{eqnarray}
we find that its matrix elements are given by
	\begin{eqnarray} \oppr{\psi_n}{\delta V}{\psi_m} = \twomatrix{0}{1}{1}{0}_{mn}, \end{eqnarray}
which can be easily verified by performing the integration explicitly in the position basis. This Hamiltonian thus generates the required rotations $R_x$ around the x-axis in Bloch space.

	The Hadamard gate $H = \frac{1}{\sqrt{2}}\twomatrix{1}{1}{1}{-1}$, for example, is given by
	\begin{eqnarray} H = e^{i\frac{\pi}{2}}R_z\left(\frac{\pi}{2}\right)R_x\left(\frac{\pi}{2}\right)R_z\left(\frac{\pi}{2}\right), \end{eqnarray}
corresponding to a free evolution for a period $\Delta t = \frac{\pi}{4\omega}$, followed by a perturbation of the potential by $\delta V$ for a period $\Delta t = \frac{\pi}{4}$ and finally followed by another period of free evolution as long as the first. We justify the neglect of the free evolution during the second time interval (the X-evolution) by choosing $\omega$ to be small (e.g.\ by choosing a small particle mass $m$). The overall phase will be relevant in interactions with other qubits, but in the Deutsch algorithm there is only one such interaction (the oracle) and hence the phase may be fixed by preparing the correct states at the right time to achieve the right phase factor for the interaction.

	We will now turn our attention to the interaction of the oracle. We will consider specific choices $f_i$. $f_0$ is trivial to implement (no specific evolution at all) and $f_3$ is simply a NOT-gate in the auxiliary qubit, which we can implement by perturbing the potential as above for an interval $\Delta t = \frac{\pi}{2}$. In neither of these cases is any interactions between the qubits required. However, for the cases of balanced functions $f_i$, i.e.\ for $f_1$ and $f_2$, the situation is different and an actual two-qubit gate needs to be constructed.
	
	We will only consider $f_2$ here. A construction of $f_1$ is easily obtainable by modifying the description here, or by using the $f_2$ evolution exactly, but placing an additional NOT-gate before and after the oracle in the data qubit. We implement the oracle by perturbing the potential by an additional term $U(x,y)$, where $x$ and $y$ denote the position coordinates of the data and auxiliary qubits respectively. We remark that if physically we have a single particle in two dimensions, such a perturbation may be easily visualised. For two particles in one-dimensional wells, this term corresponds to a (possibly quite complicated) coupling between them. We recall that the required Hamiltonian is abstractly given by
	\begin{eqnarray}\label{f2Hamiltonian} TH_{oracle\text{ }ij} &=& \notag \twomatrix{\delta_{1,f(0)}\frac{\pi}{2}(X-\mathds{1})}{0}
																					{0}{\delta_{1,f(1)}\frac{\pi}{2}(X-\mathds{1})}_{ij} \qquad\qquad 
																							\text{for } i,j\in \{1,2,3,4\}\\
																				 &\underset{f_2}{=}& \twomatrix{\frac{\pi}{2}(X-\mathds{1})}{0}{0}{0}_{ij}.
	\end{eqnarray}
Choosing for simplicity the time interval during which the gate acts to be $T=\frac{\pi}{2}$, $U(x,y)$ must have the following matrix elements:
	\begin{eqnarray}
	\notag \bra{\psi_1}_d\bra{\psi_m}_a U(x,y) \ket{\psi_n}_a\ket{\psi_1}_d &=& (X-\mathds{1})_{mn} \\
	\notag \bra{\psi_1}_d\bra{\psi_m}_a U(x,y) \ket{\psi_n}_a\ket{\psi_2}_d &=& 0 \\
	\notag \bra{\psi_2}_d\bra{\psi_m}_a U(x,y) \ket{\psi_n}_a\ket{\psi_1}_d &=& 0 \\
				 \bra{\psi_2}_d\bra{\psi_m}_a U(x,y) \ket{\psi_n}_a\ket{\psi_2}_d &=& 0  
	\end{eqnarray}
with $m,n\in\{1,2\}$. This may be achieved with a potential $U(x,y,)$ inside the well of the form
	\begin{eqnarray} U(x,y) = (A+B\cos x+Cx\cos x)\left[-\frac{9\pi^2}{16}(y-\frac{1}{2})-1\right], \end{eqnarray}
where $A,B$ and $C$ are appropriately chosen constants. Here their numerical values are $A=\frac{52}{27}$, $B=-\frac{225}{432}\pi^2$ and $C=\frac{225}{216}\pi^2$. These values are easily derivable and a brief description of how they were obtained is given in the appendix. The matrix elements
	\begin{eqnarray}\notag &&\bra{\psi_p}_d\bra{\psi_m}_a U(x,y) \ket{\psi_n}_a\ket{\psi_q}_d \\
	\notag &&\qquad = \bra{\psi_p}_d(A+B\cos x+Cx\cos x)\ket{\psi_q}_d\bra{\psi_m}_a\left[ -\frac{9\pi^2}{16}(y-\frac{1}{2})-1\right]\ket{\psi_n}_a \\
	\end{eqnarray}
then do indeed satisfy the requirements given above as may once again be verified by explicit calculation in the position basis. Of course, multiple other choices for $U(x,y)$ using linear combinations of other functions of $x$ are possible.

	We can now analyse the evolution of the combined wave function of the two qubits by modifying the potential of the well(s) in the right way for the right amount of time. By construction we obtain the evolution of the Deutsch algorithm as abstractly described in section \ref{DeutschAlgorithm}.

	The final measurement is an interaction between the data qubit and the pointer, which is described by a position variable $z$ and whose wave function starts as an initial narrow wave packet centered at $z=0$. An appropriate interaction Hamiltonian is given by
	\begin{eqnarray} \hat{H}_{meas} = a(-i\partial_x)^2(-i\partial_z) = ia\partial_x^2\partial_z. \end{eqnarray}
Its analysis is exactly analogous to previous discussions of measurement. The ensemble density simply shifts by $a\delta t\cdot n^2\pi^2$ for $n=1,2$ corresponding to the two energy states, and the projection of the trajectories onto the configuration space dimension of the pointer are all exactly parallel. 

	We see that we can fully account for the outcome of the computation by only considering the wave function up to the point of measurement and then appeal to standard PWT measurement theory. As such, we have also automatically analysed the evolution of the ensemble, at least in quantum equilibrium. A corresponding analysis for quantum non-equilibrium would possibly present interesting future work, although we can already conclude that if $\psi = 0 \rightarrow \rho = 0$ (i.e.\ the density function only has ``support'' where the pilot-wave is non-zero), then the computation works exactly the same nevertheless: The measurement outcome is deterministic in every sense (even in an SQM description) and only depends on the choice of oracle function $f$, hence \emph{any} trajectory whose corpuscle is at some time inside the configuration space area where $\psi\neq 0$ corresponds to the correct computational output.
	
	The only possible task left before us is a purely technical exercise, namely to calculate trajectories of the configuration space particle (in the $x$-$y$ subspace of configuration space) during the action of the gates. Note that the aspect most crucial for the trajectory is its projection onto the dimension corresponding to the degree of freedom of the pointer of the measurement apparatus. This we already know. The trajectories of the qubits themselves may be investigated here for completeness only. Since we are dealing with systems of spinless particles whose Hamiltonian is of the form $H_{well}=-\frac{1}{2m}\nabla^2+V$, the guidance equation is the usual
	\begin{eqnarray} \dot{\vec{x}} = \frac{1}{m}\nabla S. \end{eqnarray}
We are particularly interested in how the trajectories of the particles differ between the cases of a constant and a balanced oracle function $f$. We recall that, when entering the oracle, the data qubit is in a state 
\begin{eqnarray}\ket{+}_d = \frac{1}{\sqrt{2}}(\ket{0}_d+\ket{1}_d) = \frac{1}{\sqrt{2}}(\sqrt{2}\sin(\pi x)+\sqrt{2}\sin(2\pi x)).
\end{eqnarray}
 If $f$ is constant it will return to that state when the oracle action stops and if $f$ is balanced its final state will be \begin{eqnarray}\ket{-}_d = \frac{1}{\sqrt{2}}(\ket{0}_d-\ket{1}_d) = \frac{1}{\sqrt{2}}(\sqrt{2}\sin(\pi x)-\sqrt{2}\sin(2\pi x)).\end{eqnarray}
 The corresponding quantum equilibrium ensemble density functions in the $x$-dimension are depicted in figures \ref{plusdensity} and \ref{minusdensity} below.

\begin{figure}[H]
	\centering
		\includegraphics[width=300pt]{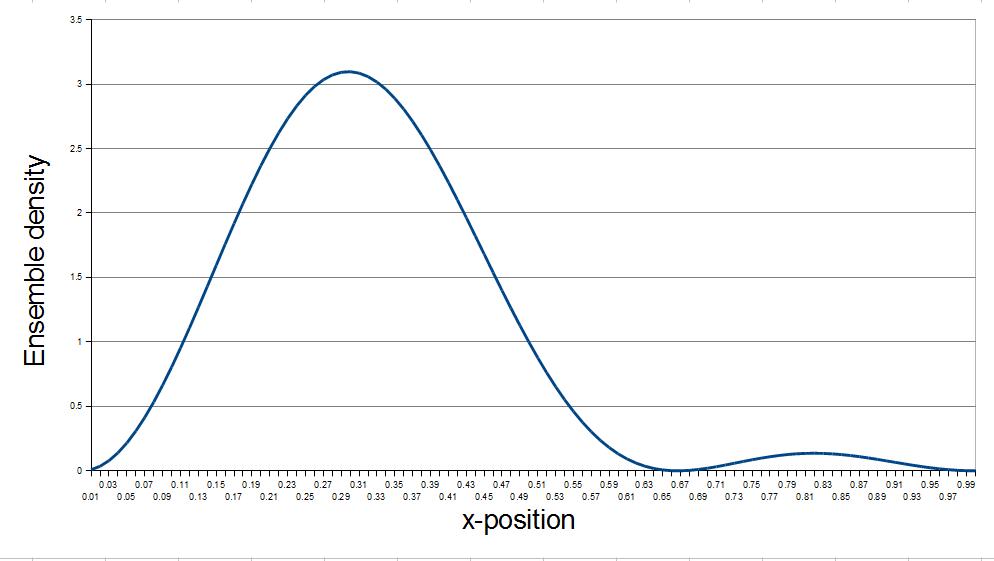}
	\caption{\small $\rho(x) = |\inpr{x}{+}|^2 = |\sin(\pi x) + \sin(2\pi x)|^2$ for an infinite well of length $L=1$.}
	\label{plusdensity}
\end{figure}
\begin{figure}[H]
	\centering
		\includegraphics[width=300pt]{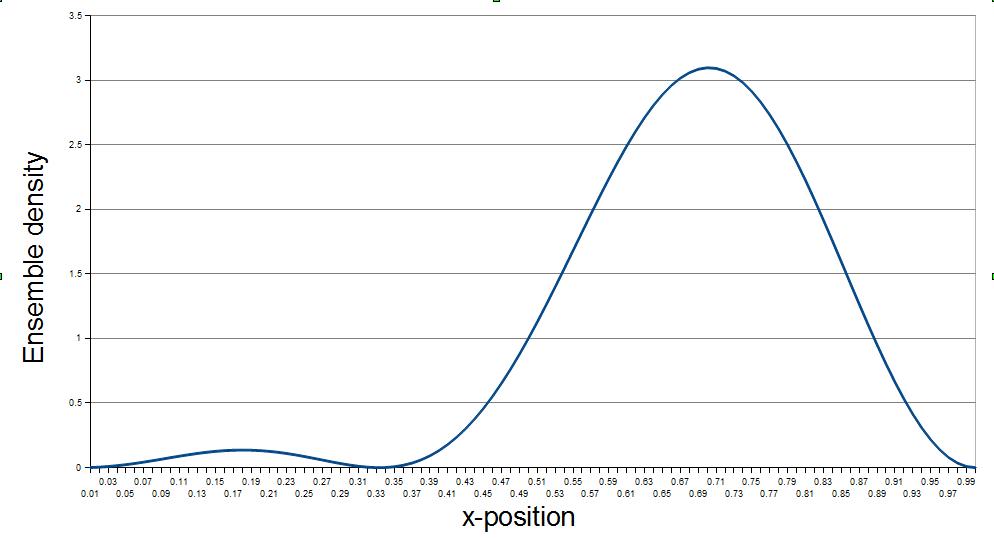}
	\caption{\small $\rho(x) = |\inpr{x}{-}|^2 = |\sin(\pi x) - \sin(2\pi x)|^2$ for an infinite well of length $L=1$.}
	\label{minusdensity}
\end{figure}

	This already highlights the key difference between the types of trajectories found in the two cases: In the constant case, the particle ensemble returns to its initial distribution of $x$-positions, in the balanced case it does not. For both a constant and a balanced oracle function, the auxiliary qubit evolves from $\ket{-}_a$ to $\ket{-}_a$ (i.e.\ its initial and final states are the same) and so the ensemble returns to its initial $y$-distribution. 
	
	We will now take a glimpse at the trajectories and indicate what a more detailed analysis (perhaps in future work) might consist of. We note that for pilot-waves in non-trivial superpositions of the two eigenfunctions $\sin(\pi x)$ and $\sin(2\pi x)$, the phase $S$ is given by the arctan of a complicated function of $x,y$ and $t$ and the trajectories cannot be found analytically, but numerical analysis called for. Here we will only consider the evolution of the configuration particle during the action of the oracle rather than the entire algorithm. We will attempt to compare the two specific cases $f_0$ and $f_2$, i.e.\ one constant and one balanced oracle function $f$. 
	
	First, however, consider the free evolution of the qubits in the absence of any gates. The phase $S$ of the states $\ket{0}$ and $\ket{1}$ is just given by $-iE_1t$ and $-iE_2t$ respectively and hence $\nabla S = 0$ in both cases. The particle is at rest. This is not the case for superpositions. A general state, up to some overall phase independent of $x$ is given by
	\begin{eqnarray} 
	\notag \ket{\psi} &=& ae^{-i\omega t}\sqrt{2}\sin(\pi x)+be^{i\omega t}\sqrt{2}\sin(2\pi x)\\
	\notag 						&=& a\cos(\omega t)\sqrt{2}\sin(\pi x)+b\cos(\omega t)\sqrt{2}\sin(2\pi x) \\
	\notag								&&\qquad	+ i[-a\sin(\omega t)\sqrt{2}\sin(\pi x)+b\sin(\omega t)\sqrt{2}\sin(2\pi x)] \\
										&=& e^{i\arctan\left( \tan(\omega t)\frac{-a\sin(\pi x)+b\sin(2\pi x)}{a\sin(\pi x)+b\sin(2\pi x)} \right)}.
	\end{eqnarray}
	
	In order to find the trajectories we have set up an algorithm computing the evolution in time steps $\delta t = 0.01$ (recall that gates act for time intervals of order $\pi$). The algorithm has been implemented in \emph{C++}. We have chosen to investigate the particular state $\ket{+}_d\ket{-}_a$, since this is the state of the qubits just prior to the action of the oracle. We have chosen to arbitrarily set the mass parameter to $m=1$. This is not a physical choice, nor is it a choice that necessarily warrants our assumption to ignore the free evolution during the action of the gates. However, it is a choice that leads to large displacements in the trajectories and as such is a choice that is useful in identifying their qualitative features. We have chosen to consider trajectories such that initially $x=y$. We have no physical motivation to do so, but it is merely a choice that provides us with some specific parameters to investigate. The trajectories are then as depicted in figures \ref{freex} and \ref{freey} below.

\begin{figure}[H]
	\centering
		\includegraphics[width=300pt]{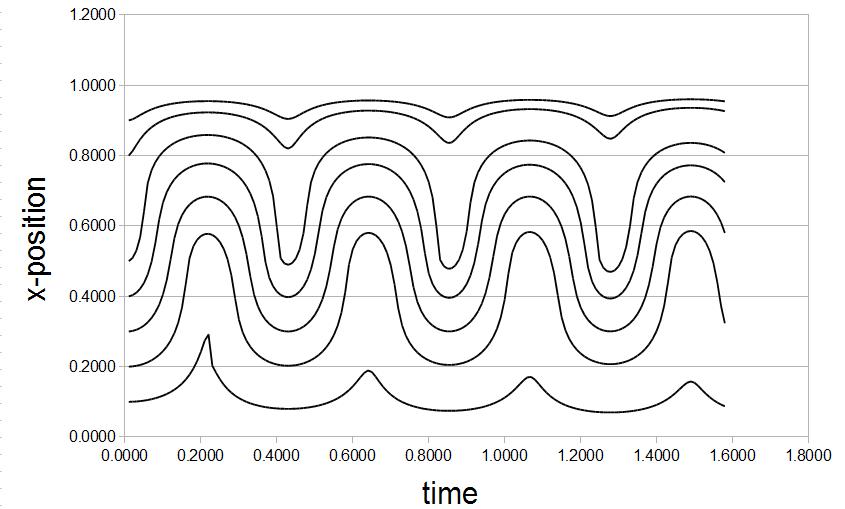}
	\caption{\small Trajectories of freely evolving qubits, $x$-position against time. Note that for initial values $x=0.6$ and $x=0.7$ the algorithm yielded large errors, which we believe are due to the inaccuracy of stepwise evolution becoming significant in areas where $|\psi|^2\rightarrow 0$. In future work, we hope to develop an improved algorithm that overcomes these difficulties.}
	\label{freex}
\end{figure}
\begin{figure}[H]
	\centering
		\includegraphics[width=300pt]{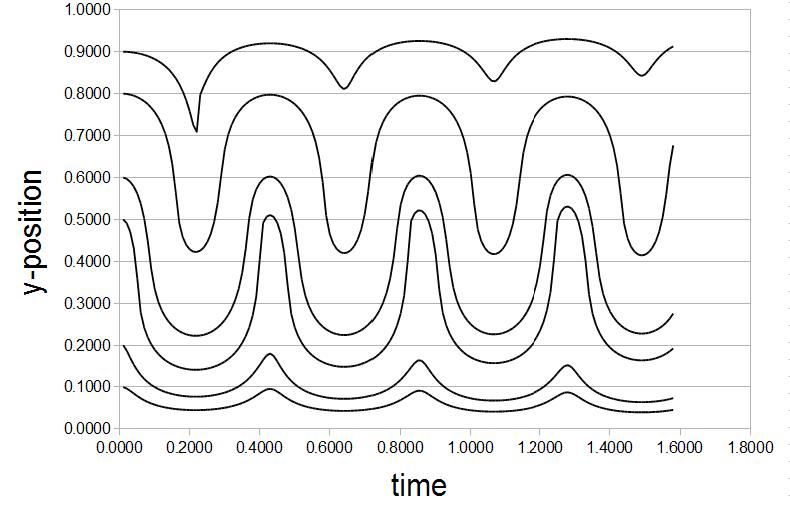}
	\caption{\small Trajectories of freely evolving qubits, $y$-position against time.}
	\label{freey}
\end{figure}

	In $x$-$y$-space these trajectories correspond to 1-dimensional oscillations. Running the algorithm for initial positions $x\neq y$ leads to similar behaviour. By modifying parameters in our program we have discovered a very significant dependence of the amplitudes of the oscillatory motion of the particles on the mass parameter $m$. For example, by increasing the mass tenfold, the amplitudes are reduced to about one hundredth as compared to the ones depicted. Future work should therefore also include a more physical parameter choice. 
	
	Let us now investigate the oracle. For the constant function $f_0$, the corpuscle remains at rest (neglecting its free evolution), from which the ensemble evolution described in figures \ref{plusdensity} and \ref{minusdensity} follows trivially. Let us contrast this to the evolution generated by the corresponding Hamiltonian, given in equation \ref{f2Hamiltonian}. One result to note immediately is that the particle does not move at all in the $y$-dimension, no matter what its initial values for $x$ and $y$ are. Thus its ensemble density evolution is again recovered naturally. The $x$-evolution is more interesting. We have computed some trajectories for typical $x$-values (i.e.\ for $x$-values with significant non-zero ensemble density) below in figure \ref{oraclex}. Note that we here chose $m=10$ since lower mass parameters resulted in erratic behaviour near the boundaries $x=0$ and $x=1$. Once more we leave the development of an improved algorithm to future work. The code for the simple algorithm used here is given in the appendix.
	
\begin{figure}[H]
	\centering
		\includegraphics[width=0.80\textwidth]{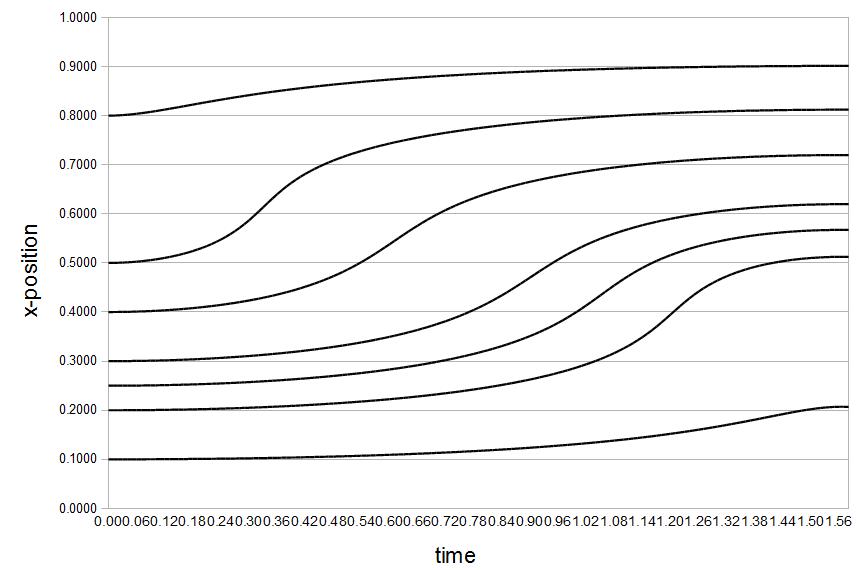}
	\caption{Trajectories in the $x$-dimension for an oracle implementing function $f_2$}
	\label{oraclex}
\end{figure}

	The trajectories are consistent with our expectation (figure \ref{plusdensity}). The ensemble density shifts in $x$-space from its initial to its final distribution in relatively simple paths. Recall that in the case of Bell's spin model no spatial motion at all was necessary (except for the pointer) in order to account for the computation of the Deutsch algorithm. It may be interesting to investigate other types of physical implementations of the circuit and identify any common properties of the trajectories, although ultimately we should not expect too much from such an exercise. Other than the theory of measurement of PWT, the evolution of the wave function alone is sufficient to explain quantum computing.

%______________________________________________________________________________________________________________________		
		
	\pagebreak		
	\section{Conclusions}\label{Conclusions}
	
	By example of the Deutsch algorithm we have seen how de Broglie-Bohm Pilot-Wave Theory is able to account for quantum computation just as well as standard quantum mechanics or many-world theories. Additionally, unlike SQM, PWT provides an explicit theory of measurement in terms of the interaction between the measurement apparatus (some system with an ``output'') and the system to be measured.
	
	In our analysis we have observed that most of the interesting physical evolution that constitutes the computation is the evolution of the wave function (the pilot-wave) rather than the motion of the configuration particle. In particular, in Bell's (somewhat oversimplified) model for spin we saw that the particle is at rest until the final measurement and only moves as the result of the interaction between one of the qubits and the pointer. 
	
	This provides strong support for Mermin's position that it is properties of wave behaviour that allow for the relative speed-up of quantum computing as compared to classical computing. Josza's view is partially compatible with this: The oracle in the Deutsch algorithm entangles two qubits, although in the algorithm itself the resultant state is a product state because of the specific choice of initial states. Entanglement clearly does play a role, although Josza's claim that entanglement is the key ingredient for quantum computing appears to overstate its significance. Either way, we might consider wave properties to be the more fundamental concept (as without it entanglement is not possible) anyway.
	
	If we however side with Goldstein et al.\ and understand the wave function to be not a physical entity but law-like, or nomological, our conclusion would have to be modified. Clearly it cannot have been the evolution of the wave function that constitutes the computation prior to the final output, since the wave function is not a real entity in this picture. But as the analysis of the Deutsch algorithm in terms of Bell's spin model has revealed, the particle itself does not have to move in any way until the final measurement. This leads to a peculiar conclusion: According to Goldstein's view, there does not have to exist any physical evolution that we could take to constitute the computation. Instead, nature happens to be such that if we set up all our gates in the right way, then the laws of motion are such that the particle's direction of motion is in a one-one correspondence with $f$ being constant and balanced. One option here is to simply accept this conclusion as it stands. Another would be to criticise Bell's and any similar toy model as unrealistic. Alternatively, we might see the argument as providing further support to the view that a real physical pilot-wave is called for.
	
	By appealing to the possibility of describing quantum computing in PWT we have also shown that Steane is right: a quantum computer really only needs one universe. However, this alone will not convince Deutsch. As stated, the interesting process in the computation is the evolution of the wave function, not the trajectory of the particle. Deutsch will therefore remind us that PWT is a many-world theory in denial as the wave function contains the structure of many worlds.\footnote{Here we are not talking about Deutsch's point concerning unoccupied grooves, which would concern parallel universes to which ours has no physical connection and as such these other universes are in no way relevant to quantum computing in our universe. Instead we are concerned with the point more accurately defended by Brown and Wallace that all the required structure is in the wave function and the corpuscle is a superfluent ``epiphenomenal pointer''.} However, he will thereby have to admit that quantum computing does not provide any further support at all towards a many-world picture, since the issue is theoretical, not phenomenological. This is not unexpected as we have established that quantum computers are only a particular type of physical system, subject to the laws of quantum mechanics. 
	
	Contrary to Deutsch's claims we have no reason to believe that the function $f(x)$ is computed for different values $x$ in parallel universes, which then interfere in the right sort of way. While it is tempting to think so from our mathematical description of the evolution, a computation is a particular kind of physical evolution with an output, but the only output is the single bit at the end encoding whether $f$ is constant or balanced. To validate the claim that $f(0)$ and $f(1)$ are computed simultaneously Deutsch would need to provide a modified algorithm that includes an output for both of these. This however is impossible with a single function call. What follows from this discussion is this: Whether MWT or PWT or some other theory is closest to the real nature of the physical world is a question that cannot be settled by appealing to quantum computing. 
	
	Instead the question is one concerning the internal consistency of these theories. We have observed that both of them leave questions that have not yet been answered satisfactorily, such as questions concerning the emergence of the physical 3-space world. It is also questions concerning emergence that appear to constitute the greatest division between PWTists and MWTists, although in most discussions this is not obvious. MWTists consider the wave function and its structure sufficient to explain our physical world. PWTists disagree and see the necessity of postulating this emergence explicitly in the form of an \emph{actual} configuration and trajectory in configuration space, thereby also providing an account for the statistical frequency of measurement outcomes that matches the Born rule of SQM.
	
	If the choice is purely between PWT and MWT in its strongest form (i.e.\ as a ``wave function only'' theory that does not fall into the trap of eigenvalue realism or premature talk of many parallel worlds), at this point our decision has to be at least to some degree dogmatic. What is required for the emergence of physical reality as we know it? Perhaps future research will allow us to provide more conclusive arguments in favour of one of the two, although, unless quantum non-equilibrium is discovered, the question will ultimately not be one that can be answered experimentally, through quantum computation or otherwise.

\section*{\large Acknowledgements}
My special thanks goes to Antony Valentini for his support in both technical and stylistic questions, as well as thought-provoking discussions. I would also like to thank various members of the Philosophy of Physics group at Oxford University for first allowing me to appreciate the conceptual intricacies of quantum physics.

%_________________________________________________________________________________
\pagebreak
\section{Appendices}

	\subsection*{Appendix A: Non-locality in PWT}
	
 PWT is a hidden variable theory and so according to Bell's theorem must be non-local if it is to reproduce the experimental predictions of SQM, which it does. In fact, PWT is very explicitly non-local. The implication is that PWT relies on a Lorentzian spacetime, that is, a spacetime with a single preferred simultaneity foliation. In fact, parametrising the evolution of the wave function in configuration space by a parameter $t$ already makes this evident. Note however that this preferred frame is not detectable (in quantum equilibrium) and so special relativity is not empirically violated.
 
 Our discussion here is bound to be somewhat artificial, given that we are considering a non-relativistic formulation of PWT. However, it is nevertheless instructive to see how non-locality arises and how locality is recovered at the phenomenological level.
 
 Suppose that we have a system of two particles $X$ and $Y$ with coordinates $\vec{x}$ and $\vec{y}$ in 3-space and masses $m_X$ and $m_Y$ respectively. The particles may be arbitrarily (lightyears etc.) far apart. Their dynamics is governed by a Hamiltonian $\hat{H}_{XY} = \hat{H}_X+\hat{H}_Y$ and their joint wave function is denoted as $\psi(\vec{x},\vec{y},t)$. For standard Hamiltonians, the equations of motion of the particles are given by
 \begin{eqnarray} \notag \dot{\vec{x}} &=& \frac{1}{m_X}\nabla_x S(\vec{x},\vec{y},t) \\
 												 \dot{\vec{y}} &=& \frac{1}{m_Y}\nabla_y S(\vec{x},\vec{y},t)	\end{eqnarray}
where $S(\vec{x},\vec{y},t)$ is the phase of $\psi(\vec{x},\vec{y},t)$. Hence the velocity of one particle is dependent on the coordinate of the other (through the phase of the joint wave function). An exception is the case when $\psi(\vec{x},\vec{y},t)$ is a product wave function (i.e.\ the particles are not entangled), that is, 
	\begin{eqnarray} \psi(\vec{x},\vec{y},t) = \psi_X(\vec{x},t)\psi_Y(\vec{y},t), \end{eqnarray}
for which the phase is just the sum of the two phases of the single-particle wave function:
	\begin{eqnarray} S(\vec{x},\vec{y},t) = S_X(\vec{x},t)+S_Y(\vec{y},t). \end{eqnarray}
Hence the equations of motion reduce to
 \begin{eqnarray} 
 \notag \dot{\vec{x}} &=& \frac{1}{m_X}\nabla_x (S_X(\vec{x},t)+S_Y(\vec{y},t)) = \frac{1}{m_X}\nabla_x S_X(\vec{x},t) \\
 				\dot{\vec{y}} &=& \frac{1}{m_Y}\nabla_y (S_X(\vec{x},t)+S_Y(\vec{y},t)) = \frac{1}{m_Y}\nabla_y S_Y(\vec{y},t)\end{eqnarray}
and full locality is recovered. 

	Note however that even in the general case of a non-separable wave function the dependence of $\dot{\vec{x}}$ on $\vec{y}$ and of $\dot{\vec{y}}$ on $\vec{x}$ only expresses a correlation, not any form of superluminal causal influence. This correlation is also manifest in SQM, for example in the violation of Bell inequalities. However, if we modify the Hamiltonian $\hat{H}_Y$ at time $t_0$, say, and let $\psi(\vec{x},\vec{y},t)$ evolve according to the Schr\"odinger equation, then for $t>t_0$ we will have causally affected $\dot{\vec{x}}$ by locally manipulating the environment of particle $Y$ only. Thus non-locality is manifested in PWT.
	
	However, in quantum equilibrium this effect is undetectable since the marginal ensemble density 
	\begin{eqnarray} \rho_X(\vec{x},t) = \int d^3y \rho(\vec{x},\vec{y},t) = \int d^3y |\psi(\vec{x},\vec{y},t)|^2 \end{eqnarray}
is independent of the evolution resulting from a purely local manipulation of $\hat{H}_Y$. Work by Valentini\footnote{1991, \cite{ValentiniHTheorem}} suggests that this is a special case and in general for non-equilibrium ($\rho\neq |\psi|^2$) the superluminal causal influence may be detectable. In other words, quantum equilibrium conceals the non-local effects and thus makes instantaneous signalling impossible. In non-equilibrium then the preferred spacetime foliation (i.e.\ the preferred frame) would be detectable, in direct violation with special relativity. This may seem problematic but fundamentally is not: Note that we have good reason to believe that our environment is in or at least very close to quantum equilibrium and so in the domain to which we have experimental access the phenomenology of special relativity is recovered. We have no \emph{a priori} reason to believe that special relativity must be true, but our faith in it, just like in the case of quantum theory, is based on experimental corroboration. If, as Valentini suggests,\footnote{2007, \cite{ValentiniAstroTests}} we may some day discover relic particles that have decoupled very early after the Big Bang and are in non-equilibrium then we might also expect empirical violations of special relativity (and thus also of Lorentzian special relativity). Currently the existence of such matter is however purely speculative.\\

\subsection*{Appendix B: Some common objections to pilot-wave theory}

\textbf{PWT is empirically indistinguishable from SQM, hence it is not really a different theory at all.} This objection clearly presupposes a positivist point of view. Unlike SQM, PWT has real trajectories, definite particle positions at all times, no arbitrary distinction between (classical?) apparatuses and quantum systems and so on. In this sense it is fundamentally different. Positivists might understand PWT as a mere reformulation, but that by itself is certainly not reason enough to reject it either. Also even granting the positivist point of view, we could turn the argument around and criticise SQM on the basis that it makes the same predictions as PWT! Historical priority is harldy on the side of SQM.\footnote{See Bacciagaluppi \& Valentini, 2009, \cite{BacciagaluppiValentini2009}}
	
	\textbf{By the principle of parsimony (Ockham's razor) PWT loses out to SQM as it requires the postulation of two entities rather than just one: the pilot-wave and the particle.} Two points can be made here: (1) While ontological simplicity certainly seems like a desirable quality of a theory (in fact, it is the cause for a whole quest in physics called unification), there is no \emph{a priori} reason to suppose that out of any two theories the one which postulates fewer entities must automatically be the ``truer'' one. (2) SQM does not merely postulate a single entity either. It includes classical objects (measurement devices) in its fundamental ontology (even though it is not explicit about it). There may be some truth in saying that SQM loses out to Everettian Many-Worlds Theory, if only the number of independent types of entities that are postulated is considered. Here we are considering a MWT where only the wave function is postulated and all structures (worlds etc.) are emergent from it. The ontology of other variants of MWT which postulate many worlds (or even ``many minds'') independently is less acceptable as these postulated entities seem somewhat ad hoc. Unsurprisingly such variants receive little support compared to pure wave function MWT in today's physics community and neither were they, it seems, Everett's original intention.
	
	\textbf{The pilot-wave affects the particle, but the particle does not affect the pilot-wave. This violates the action-reaction principle.} This objection exists in many variations, all of which presuppose that a one-way action between entities is \emph{a priori} impossible. What principle exactly is this to which these critics are referring? Certainly not Newton's third law, which is a principle concerning forces between classical objects. SQM happens to have no ``one-way'' action either, but why should this be imperative to any quantum theory? Another response based on the claim that the first-order nature of PWT means that it does not \emph{need} any kind of action-reaction principle in order to display desired symmetries is briefly mentioned in D\"urr, Goldstein and Zangh\`i.\footnote{1997, \cite{DuerrGoldsteinZanghi1997}} Their preferred answer to this criticism is however to regard the pilot-wave not as real but as nomological, or lawlike, an idea we have criticised above.
	
	\textbf{PWT is fundamentally non-local and hence violates special relativity.} We already treated this is to some extent in appendix A. In particular, two points come to mind: (1) In quantum equilibrium this non-locality does not allow superluminal signalling. While it does determine a preferred foliation of space-time into simultaneity planes, this foliation is undetectable. Our theory of relativity must therefore be Lorentzian, but this does not conflict with any empirical results. That superluminal signalling may be possible in quantun non-equilibrium is not a problem since quantum non-equilibrium has not been observed. (2) Any quantum theory empirically agreeing with SQM (including, therefore, SQM itself) violates Bell inequalities. Hence any such theory is non-local at the statistical level, even if they are not as explicit about it as PWT. In SQM, the instantaneous collapse of the wave function determines a preferred foliation, too. Ultimately, however, we should consider this question in the context of fully relativistic quantum theories, such as relativistic field theory and its PWT equivalent.
	
	\textbf{A generalisation of PWT to an equivalent of quantum field theory is impossible.} This is simply not true. There is still a considerable amount of work to be done in this area but there certainly are models that account for the results of standard QFT. One successful approach is a model proposed by Struyve and Westman\footnote{2006, \cite{StruyveWestman2006}} in which only bosonic degrees of freedom are described as ``beables'' (i.e.\ have a role analogous to the configuration particle in the non-relativistic case). While this might seem odd, note that our discussion of spin has already provided us with an example of accounting for phenomena without ascribing any ``beable''-status to them: spin is a property of the wave function only.
	
	\textbf{There is no reason the world should be in quantum equilibrium. To say that it is, is an unwarranted ad hoc assumption designed to reproduce the predictions of SQM.} For a long time there was much debate over the status of the ``equilibrium hypothesis'', especially the notion of law-like boundary conditions. Stated differently, the criticism is that the wave function acquires two logically independent meanings: as a guiding wave and as a probability density distribution. However, recently two other lines of argument have emerged to justify the assumption of quantum equilibrium: (1) Valentini's\footnote{Valentini 1991, \cite{ValentiniHTheorem}, Valentini \& Westman 2005, \cite{ValentiniWestman2005}} dynamical explanation of quantum equilibrium, showing that under certain conditions system in non-equilibrium approach equilibrium rapidly and (2) typicality arguments derived from considerations regarding the wave function of the universe and how to assign wave functions to its subsystems, proposed by D\"urr, Goldstein and Zangh\`i.\footnote{1992, \cite{DuerrGoldsteinZanghi1992}}\\

\subsection*{Appendix C: Derivation of the numerical values in potential for the infinite-well model}

	We note the standard integrals
\begin{eqnarray}
\notag \int_0^1 dx \sin^2(\pi x) &=& \frac{1}{2} \\
\notag \int_0^1 dx x\sin^2(\pi x) &=& \frac{1}{4} \\
\notag \int_0^1 dx \sin^2(\pi x)\cos(\pi x) &=& 0 \\
\notag \int_0^1 dx x\sin^2(\pi x)\cos(\pi x) &=& -\frac{4}{9\pi^2} \\
\notag \int_0^1 dx \sin^2(\pi x)\cos^2(\pi x) &=& \frac{1}{8} \\
\notag \int_0^1 dx x\sin^2(\pi x)\cos^2(\pi x) &=& \frac{1}{16} \\
\notag \int_0^1 dx \sin^2(\pi x)\cos^3(\pi x) &=& 0 \\
			 \int_0^1 dx x\sin^2(\pi x)\cos^3(\pi x) &=& -\frac{52}{225\pi^2}. 
\end{eqnarray}
			 
	We wish to consider the matrix elements $u_{pq}=\oppr{\psi_p}{u(x)}{\psi_q}$ of possible potentials $u(x)$. In the position basis these are given by
	\begin{eqnarray}
	\notag u_{11} &=& \int_0^1 dx \sin(\pi x) u(x) \sin(\pi x) = \int_0^1 dx \sin^2(\pi x) u(x)\\
	\notag u_{12} &=& \int_0^1 dx \sin(\pi x) u(x) \sin(2\pi x) = 2\int_0^1 dx \sin^2(\pi x)\cos(\pi x) u(x) \\
	\notag u_{21} &=& \int_0^1 dx \sin(2\pi x) u(x) \sin(\pi x) = 2\int_0^1 dx \sin^2(\pi x)\cos(\pi x) u(x) = u_{12} \\
	\notag u_{22} &=& \int_0^1 dx \sin(2\pi x) u(x) \sin(2\pi x) = 4 \int_0^1 dx \sin^2(\pi x)\cos^2(\pi x) u(x).\\
	\end{eqnarray}
	
	We calculate the matrix elements for some likely candidates $u(x)$ by evaluating the corresponding integrals:
\begin{table}[!th] 
\begin{tabular}[c]{| c || c | c | c | c |}
	\hline
	$u(x)$ 	& $u_{11}$ & $u_{12} = u_{21}$ & $u_{22}$	 \\
	\hline\hline
	$1$ &	$1$	&	$0$	&	$1$	\\ 
	\hline
	$x$ & $\frac{1}{2}$ & $-\frac{16}{9\pi^2}$ & $\frac{1}{2}$	\\ 
	\hline
	$\cos x$ & $0$ & $\frac{1}{2}$ & $0$	\\
	\hline
	$x\cos x$ & $-\frac{8}{9\pi^2}$ & $\frac{1}{4}$ & $-\frac{416}{225\pi^2}$\\
\hline
\end{tabular} 
\end{table}

	Now we only need to find a linear combination of $1,x,\cos x$ and $x\cos x$ such that the total matrix elements match the matrix for the data qubit as required in the Hamiltonian. In this case the matrix is $\twomatrix{1}{0}{0}{0}$ since the Hamiltonian is $\twomatrix{X-\mathds{1}}{0}{0}{0} = \twomatrix{1}{0}{0}{0}\otimes (X-\mathds{1})$. This may be achieved by only using a combination of three of the four candidates in the table, such as $1,\cos x$ and $x\cos x$. We hence need to solve the simultaneous equations
	\begin{eqnarray}
	\notag A-\frac{8}{9\pi^2}C &=& 1 \\
	\notag \frac{1}{2}B+\frac{1}{4}C &=& 0 \\
				 A-\frac{416}{225\pi^2}C &=& 0.
	\end{eqnarray}
The solutions are the numerical values given in section \ref{wellmodel}, namely
	\begin{eqnarray} A = \frac{52}{27}, \qquad B = -\frac{225}{432}\pi^2, \qquad C = \frac{225}{216}\pi^2. \end{eqnarray}

\subsection*{Appendix D: \emph{C++} implementation of algorithm to calculate trajectories inside the oracle numerically}

\scriptsize
\begin{verbatim}
#include <iostream>
#include <fstream>
#include <cmath>
#include <complex>

using namespace std;

//----------Global variables----------
    const double pi = 3.141592653589793;

    double m = 10;
    double delta_t = 0.01;   //time step size

    complex<double>  psi;

//----------Evolution of coefficients----------
complex<double> evolve_a(complex<double> a, complex<double> b) {
    a = a + complex<double>(0,1)*(a-b)*delta_t;
    return a;
    }

complex<double> evolve_b(complex<double> a, complex<double> b) {
    b = b - complex<double>(0,1)*(a-b)*delta_t;
    return b;
    }

//----------Calculation of Psi-value---------
complex<double> Psi(double x, double y, complex<double> a, complex<double> b, 
		complex<double> c, complex<double> d) {
    psi = complex<double>(2,0)*(a*sin(pi*x)*sin(pi*y) + b*sin(pi*x)*sin(2*pi*y) 
    													+ c*sin(2*pi*x)*sin(pi*y) + d*sin(2*pi*x)*sin(2*pi*y));
    return psi;
    }


int main()
{
//----------Initialise variables and constants----------
    double x, y, t;               //x-position, y-position, time

    double dSdx, dSdy;            //spatial derivatives of phase

    double delta_x = 0.0001;      //precision for numerical calculation of gradient in x
    double delta_y = 0.0001;      //precision for numerical calculation of gradient in y


    complex<double> a1, b1; 			//temporary value holders

    //initial state: |+-> = 1/2[|00>-|01>+|10>-|11>]
    complex<double> a = 0.5;
    complex<double> b = -0.5;
    complex<double> c = 0.5;
    complex<double> d = -0.5;


//----------Request input for initial positions----------
    cout << "Initial x-value of corpuscle, 0 <= x <= 1: ";
    cin >> x;

    cout << "\n Initial y-value of corpuscle, 0 <= y <= 1: ";
    cin >> y;

    t = 0;


//----------Create output files and write initial values----------
    ofstream tfile ("tfile.txt");
    ofstream xfile ("xfile.txt");
    ofstream yfile ("yfile.txt");

    ofstream psifile ("psifile.txt");       //control files
    ofstream dSdxfile ("dSdxfile.txt");
    ofstream dSdyfile ("dSdyfile.txt");
    ofstream Sfile ("Sfile.txt");
    ofstream afile ("afile.txt");
    ofstream bfile ("bfile.txt");

    tfile << t << "\n";
    xfile << x << "\n";
    yfile << y << "\n";


//----------Evolve one step according to guidance equation and write to file----------
    for (t=0; t<(pi/2); t+=delta_t) {

        if(x+0.5*delta_x > 1) {delta_x = 0.5*(1-x);} //to avoid calculation with ill-defined phase values.
        if(x-0.5*delta_x < 0) {delta_x = 0.5*x;}
        if(y+0.5*delta_y > 1) {delta_y = 0.5*(1-y);}
        if(y-0.5*delta_y < 0) {delta_y = 0.5*y;}

        a1 = evolve_a(a,b);
        b1 = evolve_b(a,b);
        a = a1; 
        b = b1;

        psi = complex<double>(2,0)*(a*sin(pi*x)*sin(pi*y) + b*sin(pi*x)*sin(2*pi*y) 
        													+ c*sin(2*pi*x)*sin(pi*y) + d*sin(2*pi*x)*sin(2*pi*y));

        dSdx = ( arg(Psi(x+0.5*delta_x,y,a,b,c,d)) - arg(Psi(x-0.5*delta_x,y,a,b,c,d)) ) / delta_x;   
        if(dSdx*delta_x > pi) {dSdx = (dSdx*delta_x - 2*pi) / delta_x;}  //correcting for phase jumps -pi to +pi
        if(dSdx*delta_x < -pi) {dSdx = (dSdx*delta_x + 2*pi) / delta_x;} //correcting for phase jumps +pi to -pi

        dSdy = ( arg(Psi(x,y+0.5*delta_y,a,b,c,d)) - arg(Psi(x,y-0.5*delta_y,a,b,c,d)) ) / delta_y;   
        if(dSdy*delta_y > pi) {dSdy = (dSdy*delta_y - 2*pi) / delta_y;}  //correcting for phase jumps -pi to +pi
        if(dSdy*delta_y < -pi) {dSdy = (dSdy*delta_y + 2*pi) / delta_y;} //correcting for phase jumps +pi to -pi

        psi = Psi(x,y,a,b,c,d);

        x = x + delta_t * dSdx/m;    //stepwise x-evolution
        y = y + delta_t * dSdy/m;    //stepwise y-evolution

        tfile << t+delta_t << "\n";  

        Sfile << arg(psi) << "\n";  //control files
        dSdxfile << dSdx << "\n";
        dSdyfile << dSdy << "\n";
        psifile << psi << "\n";
        afile << a << "\n";
         bfile << b << "\n";

        xfile << x << "\n";
        yfile << y << "\n";
        }

    return 0;
}


\end{verbatim}
\normalsize

%_________________________________________________________________________________

\pagebreak

%_________________________________________________________________________________

\end{document}